\documentclass[11pt,onecolumn]{article}
\usepackage[utf8]{inputenc}
\usepackage[margin=1in]{geometry}
\usepackage[square,numbers,comma,sort&compress]{natbib}
\usepackage{textcomp,graphicx,bm,amsmath,xcolor}
\usepackage{amssymb,color,caption,subcaption,setspace}
\usepackage[square,numbers,comma,sort&compress]{natbib}

\usepackage{setspace}   
\usepackage{lipsum} 
\usepackage{wrapfig}
\usepackage[symbol]{footmisc}
\usepackage{authblk}
\usepackage{float}

\date{}
\begin{document} 


\title{Semiconductor thermal and electrical properties decoupled by localized phonon resonances\footnote{Contribution of an agency of the U.S. government; not subject to copyright.}}

\newcounter{repeatsym}
\setcounter{repeatsym}{3}
\author[a]{Bryan T. Spann\thanks{Current affiliation: Lockheed Martin Space, Advanced Technology Center, Louisville, CO 80027}\footnote{Equal contributors}}
\author[a]{Joel C. Weber$^{\fnsymbol{repeatsym}}$}
\author[a]{Matt D. Brubaker} 
\author[a]{Todd E. Harvey}
\author[b]{Lina Yang}
\author[c]{Hossein Honarvar\footnote{Current affiliation: ConcertAI, Cambridge, MA 02138, USA}} 
\author[c]{Chia-Nien Tsai} 
\author[d]{Andrew C. Treglia\footnote{Current affiliation: Department of Physics, Colorado State University,  Fort Collins CO  80523}}
\author[d]{M. Lee}
\author[c,d]{Mahmoud I. Hussein\footnote{Corresponding author: Mahmoud I. Hussein (mih@colorado.edu)}} 
\author[a]{Kris A. Bertness\footnote{Corresponding author: Kris A. Bertness (kris.bertness@nist.gov)}} 

\affil[a]{Physical Measurement Laboratory, National Institute of Standards and Technology (NIST), Boulder, CO 80302 USA}
\affil[b]{School of Aerospace Engineering, Beijing Institute of Technology, Beijing 100081, China}
\affil[c]{Ann and H.J. Smead Department of Aerospace Engineering Sciences, University of Colorado Boulder, Boulder, Colorado 80303, USA}
\affil[d]{Department of Physics, University of Colorado Boulder, Boulder, Colorado 80302, USA}
\maketitle

\renewcommand\Affilfont{\itshape\small}

\begin{abstract}
Thermoelectric materials convert heat into electricity through thermally driven charge transport in solids, or vice versa for cooling. To be competitive with conventional energy-generation technologies, a thermoelectric material must possess the properties of both an electrical conductor and a thermal insulator.~However, these properties are normally mutually exclusive because of the interconnection of the scattering mechanisms for charge carriers and phonons. Recent theoretical investigations on sub-device scales have revealed that silicon membranes covered by nanopillars exhibit a multitude of local phonon resonances, spanning the full spectrum, that couple with the heat-carrying phonons in the membrane and collectively cause a reduction in the in-plane thermal conductivity$-$while, in principle, not affecting the electrical properties because the nanopillars are external to the pathway of voltage generation and charge transport. Here this effect is demonstrated experimentally for the first time by investigating device-scale suspended silicon membranes with GaN nanopillars grown on the surface. The nanopillars cause up to 21~\% reduction in the thermal conductivity while the electrical conductivity and the Seebeck coefficient remain unaffected, thus demonstrating an unprecedented decoupling in the semiconductor's thermoelectric properties.~The measured thermal conductivity behavior for coalesced nanopillars and corresponding lattice-dynamics calculations provide further evidence that the reductions are mechanistically tied to the phonon resonances. This finding breaks a longstanding trade-off between competing properties in thermoelectricity and paves the way for engineered high-efficiency solid-state energy recovery and cooling. 
\end{abstract}

\section{Introduction}
\begin{figure} [h!]
\centering
\includegraphics[scale=1]{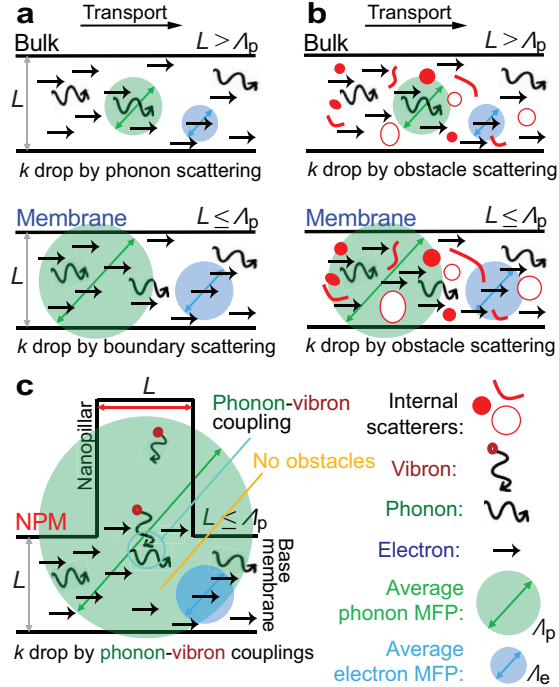}
\caption{\textbf{Prime mechanisms of thermal conductivity reduction in a semiconductor for TE conversion.} (a) bulk and reduced-dimension configurations where the key factors for $k$~reduction are phonon-phonon scattering (top), and phonon confinement and scattering off rough surfaces (bottom), respectively; (b) corresponding configurations where optimized internal scattering obstacles such as holes, inclusions, and interfaces dominate the scattering (contemporary approach); (c) NPM configuration in the form of a nanopillared membrane where the prime mechanism of $k$ reduction is resonance hybridization and the resulting phonon group velocity reductions and mode localizations (current approach).}
\label{fig:Fig1}
\end{figure}

\indent Thermoelectric (TE) materials enable electrical power generation, refrigeration, and heating, all in the solid state.~Since no moving mechanical components, fluid systems, or chemical reactions are involved, TE devices provide good reliability, stability, and overall practicality~\cite{Rowe1995}.~On the other hand, their low efficiency, around 3-6~\% in commercial devices, is a significant obstacle that impedes competitive wide-scale use as a replacement to traditional electrical power generation and fluid-based refrigeration/heat pump technologies~\cite{Yan2021}.~The performance of a TE material under a temperature gradient is based on a well-defined figure of merit, $ZT=[(S)^2\sigma/k)]T$, where $S$ is the Seebeck coefficient, $\sigma$ is the electrical conductivity, $k$ is the thermal conductivity, and $T$ is the average temperature between the hot and cold sides of the material.~The main challenge facing TE material performance is the tight coupling between these properties among nearly all classes of inorganic and organic materials; in particular, there is an inherent trade-off between exhibiting a low $k$ while simultaneously possessing a high $\sigma$ and a high $S-$a combination of attributes needed for a significant increase in $ZT$.~This trade-off has stood as a key limitation to TE technological development and proliferation since the early days of discovery of the Seebeck~\cite{Seebeck1826} and Peltier~\cite{Peltier1834} effects close to two hundred years ago~\cite{Beretta2019}.\\
\indent Increase of $ZT$ by thermal conductivity reduction is a widely pursued strategy.~Central to this path are phonon confinement~\cite{Balandin_1998} and the key scattering mechanisms available for impeding phonon transport; these include phonon-phonon scattering (which increases with temperature)~\cite{Berman1978}, boundary scattering (such as rough boundaries)~\cite{Donadio_2009,He_2012,neogi2015tuning}, and scattering by impurities and internal barriers~\cite{Biswas_2012} (Fig.~\ref{fig:Fig1}a,b).~With the advent of nanotechnology, nanostructuring has enabled precise access and control of the internal microstructure of existing materials, especially semiconductors.~A prevailing approach is the introduction of obstacles, such as holes, inclusions, and interfaces, within the interior of the TE medium to enhance phonon scattering and reduce $k$~\cite{Snyder2008,Vineis2010}.~However, in addition to scattering the phonons, the motion of charge carriers is likely to be impeded by the same obstacles.~While it is possible to tune the separation distances between scattering centers to selectively scatter phonons with longer mean free paths (MFPs) and minimize electron/hole scattering at shorter MFPs,~the problem remains constrained because the allowable range of selective scattering is limited by the inherent overlap in the intrinsic phonon and charge carrier MFP distributions of the material~\cite{ Hochbaum2008,Boukai2008}.~This in turn negates the possibility of a true decoupling of the phononic and electronic properties and subsequent realization of a substantial increase in $ZT$.

\section{Results and Discussion}
Departing from this constraining trade-off strategy, here we demonstrate decoupling of the thermal conductivity reduction from the remaining TE properties by mechanistic means.~This is done by forming a thin, suspended membrane of Si with a random arrangement of closely-packed GaN nanopillars standing on its top surface, i.e., exterior to the membrane nominal cross-section.~The membrane thickness and nanopillar spacing are selected to fall within the range of the phonon MFP distribution for Si, which is estimated to average at around 200-300 nm at room temperature~\cite{Esfarjani2011, JuGoodson1999}.~Similarly, the nanopillar feature sizes, i.e., the height and width, are selected to fall within the MFP distribution of GaN; \textit{ab initio} calculations predict that most of the thermal conductivity of GaN arises from phonons with MFPs greater than 200 nm~\cite{Garg_2018}.~The dominant portion of the heat transported in this nanostructured material is carried by traveling phonons, because generally the electronic contribution to the thermal conductivity in silicon is negligible even with heavy doping~\cite{Ohishi2015, Kim2015}.~The atoms making up the nanopillars, on the other hand, generate vibrons, or wavenumber-independent phonon resonances.~These two types of waves, the travelling and the standing, couple (Fig.~\ref{fig:Fig1}c) and cause a substantial portion of the energy of the heat-carrying phonons to modally localize in the nanopillars.~In addition, the coupling causes the base-membrane phonon group velocities to drop significantly.~These two effects lead to a reduction in the lattice thermal conductivity along the membrane portion and form the basis of the notion of a \textit{nanophononic metamaterial} (NPM)~\cite{DavisHussein2014,Wei2015,Honarvar2016a,Xiong2016,Honarvar2018,Hussein_2018,Hussein2020}.~This mechanism of phonon hybridizations and resonance localizations$-$which, in principle, takes place across the full phonon spectrum$-$is independent of the mechanisms of voltage generation and electrical charge transport and is therefore not expected to affect the Seebeck coefficient or the electrical conductivity. \\
\begin{figure*}[t!]
\centering
\includegraphics[scale=1]{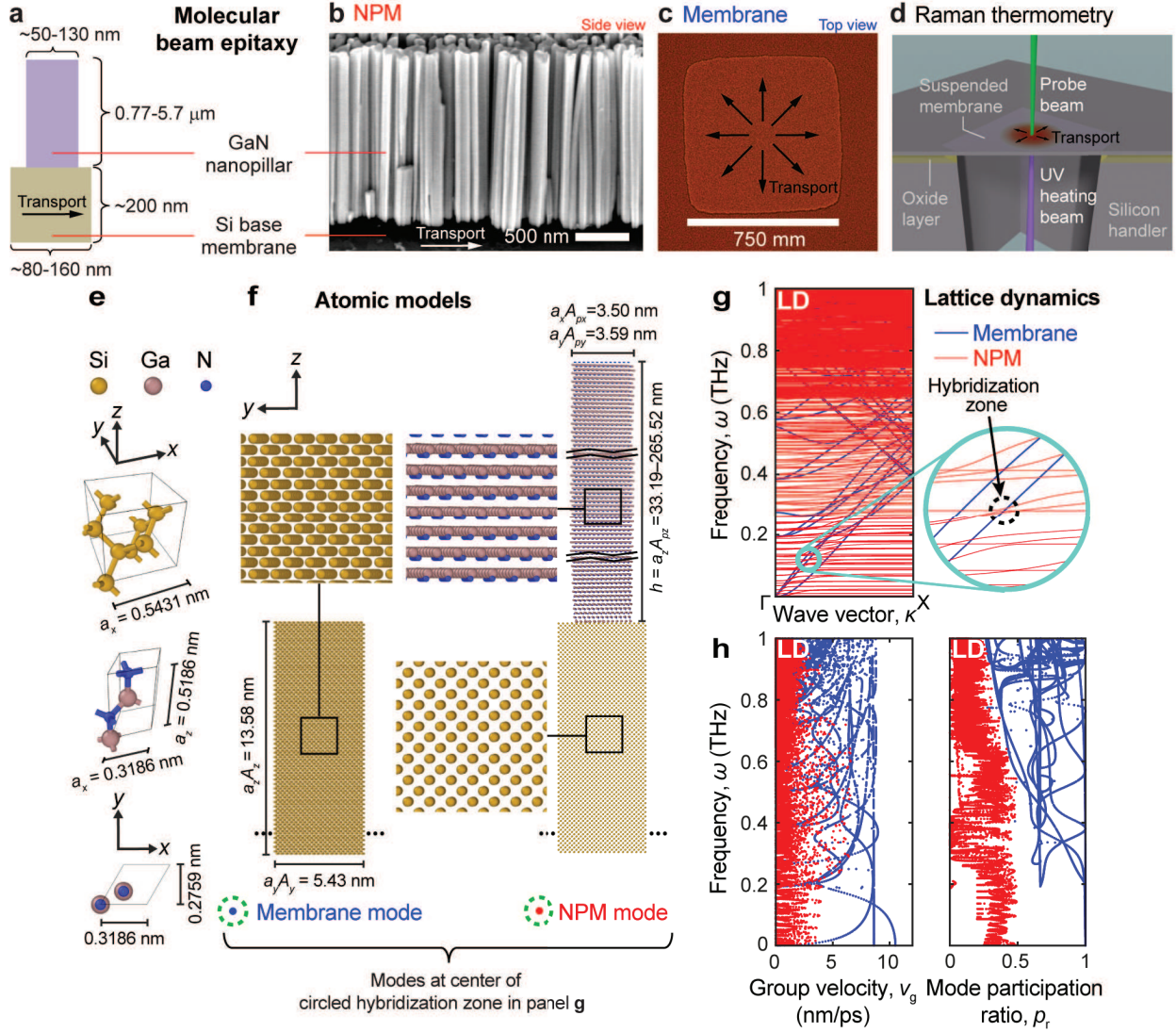}
\caption{\textbf{Nanofabricated samples of GaN-on-Si NPMs and corresponding lattice dynamics properties} (a) Schematic of the NPM unit cell, (b) SEM image of GaN nanopillars on a Si membrane.~(c) Optical microscope image of a suspended membrane, which appears lighter due to its partial transparency in the visible spectrum.~The nanopillars produced a textured appearance, and (d) schematic of the Raman thermometry measurement geometry.~(e) Conventional unit cell of Si; primitive unit cell of GaN;~(f) atomic displacements for a bare membrane mode indicating intense motion (left) and for a corresponding NPM mode indicating localized motion in the nanopillars and minimal motion in the base membrane (right). (g) Phonon band structure, and (h) group velocity (left) and mode participation (right) distributions of Si membrane with (red) or without (blue) GaN nanopillars standing on the surface.~The resonance hybridization (phonon-vibron) coupling phenomenon is illustrated in the circular inset in (g).}
\label{fig:Fig2}
\end{figure*} 
\indent Previous theoretical investigations using molecular dynamics (MD) simulations have shown the presence of phonon-vibron couplings~\cite{Honarvar2016a} and predicted up to two orders of magnitude reduction in the thermal conductivity~\cite{Honarvar2018}.~However, these studies were done on model sizes on the order of 10-20 nm for the base membrane thickness due to computational limitations.~Small nanostructures are more amenable to coherent wave effects; the key challenge is sustaining these effects at larger scales closer to the average MFP~\cite{Tang2010,Yu2010,Hussein2020}.~Here, we demonstrate the first experimental evidence for both the thermal conductivity reduction by nanoresonators (designated as the~\textit{NPM effect}~\cite{DavisHussein2014}) and the decoupling with the electrical properties, $S$ and $\sigma$.~Importantly, this demonstration is accomplished with device-scale structures, with the smallest dimension being the membrane thickness of 200 nm.~The thermal conductivity along the base membranes decreases as the nanopillars increase in height, consistent with NPM theory~\cite{Honarvar2018}.~Electrical conductivity and Seebeck coefficient measurements on the same structures show that the nanopillars do not degrade the electrical properties.~We also show that the behavior of the thermal conductivity for coalesced nanopillars provides evidence that the reductions are primarily due to phonon resonances and not boundary scattering. 

\indent The thermal conductivity test structures are illustrated in Fig.~\ref{fig:Fig2}.~The GaN nanopillars were grown on silicon-on-insulator (SOI) substrates via plasma-assisted molecular-beam epitaxy (MBE), see Fig.~\ref{fig:Fig2}b.~The GaN nanopillars formed spontaneously at high growth temperature and high N:Ga flux ratio~\cite{bertness2006}.~Specimen sets with varying nanopillar height were grown with the expectation that taller, more massive nanopillars would produce more vibrons and therefore a greater reduction in the thermal conductivity~\cite{Honarvar2018}.~The samples are of two types, Set A in which GaN nanopillar growth was initiated directly on the Si after a brief nitridation step, and Set B in which a 8-nm AlN buffer was grown prior to nanopillar growth.~As described in more detail in the Appendix, the sets differ in their electrical conductivity variation with nanopillar height because of different degrees of diffusion of Ga and Al into the membrane during high-temperature nanopillar growth.~Set A displays an increase in electrical conductivity as a function of nanopillar height, while Set B displays approximately constant electrical conductivity.~Suspended membranes were formed by etching from the backside of the substrate to the buried oxide layer, then removing the oxide layer.~The as-purchased SOI device layer thickness of 200 nm thus becomes the final membrane thickness. We note that SOI substrates with such thin device layers are only available with very light p-type doping, and therefore the electrical conductivity of these structures is not optimal for high $ZT$. This limitation is not fundamental and does not interfere with the novelty of the mechanism for thermal conductivity reduction.~The membranes were heated with a strongly absorbed ultraviolet (UV) laser beam incident from the unpatterned lower side, and the specimen temperature was measured at the center of the hot spot using a green laser beam incident from the top side (Fig.~\ref{fig:Fig2}d).~The temperature was determined by the shift in frequency of the Si Raman peak appearing near 520 cm$^{-1}$ at room temperature.~Raman thermometry is a non-contact technique that has been widely used to measure the thermal conductivity of a variety of thin membranes~\cite{Cai2010,Luo2014,luo2015, chavez2014,neogi2015tuning}.\\
\begin{figure*}[t!]
\centering
\includegraphics[scale=1]{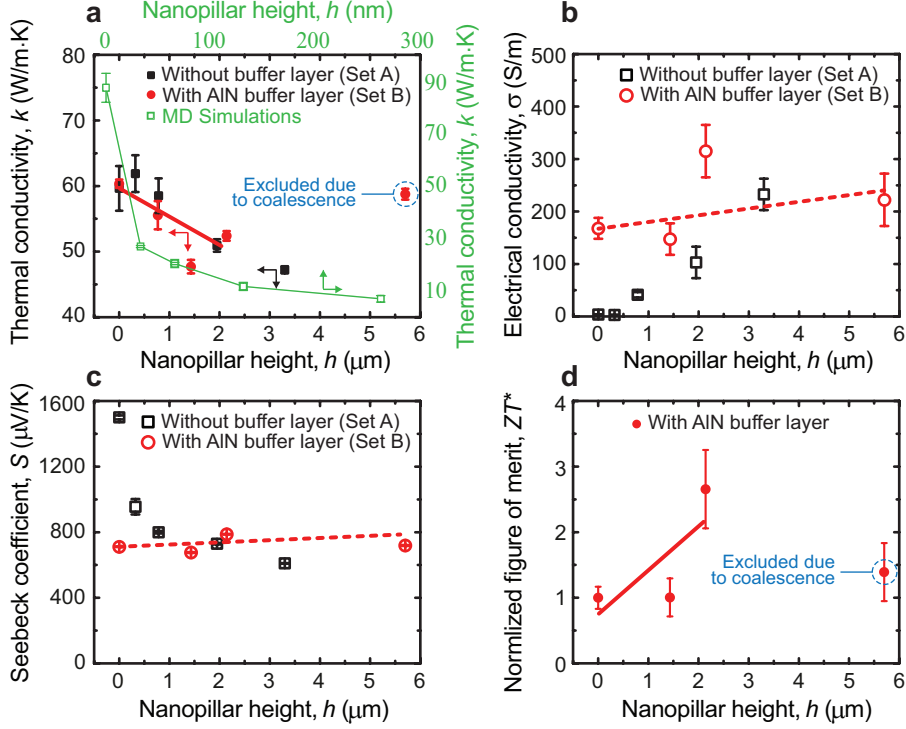}
\caption{\textbf{Measurements of TE properties of GaN-on-Si NPMs with varying nanopillar height} (a) Thermal conductivity, (b) electrical conductivity, (c) Seebeck coefficient, and (d) $ZT^*$ figure of merit normalized with respect to bare membrane value.~In (a), thermal conductivity predictions by MD simulations for smaller (by a factor of $\sim$15) but proportionally-sized models are shown in green; arrows point to relevant axes.~The AlN buffer layer (Set B) minimized diffusion of GaN into the Si membrane that dominated electrical properties in Set A. Data points circled in blue represent samples with coalesced nanopillars and were excluded from the curve fittings.~Solid (dashed) curves represent phenomenon influenced (uninfluenced) by the nanopillar vibrons.}
\label{fig:Fig3}
\end{figure*}
\begin{figure*}[t!]
\centering
\includegraphics[scale=1]{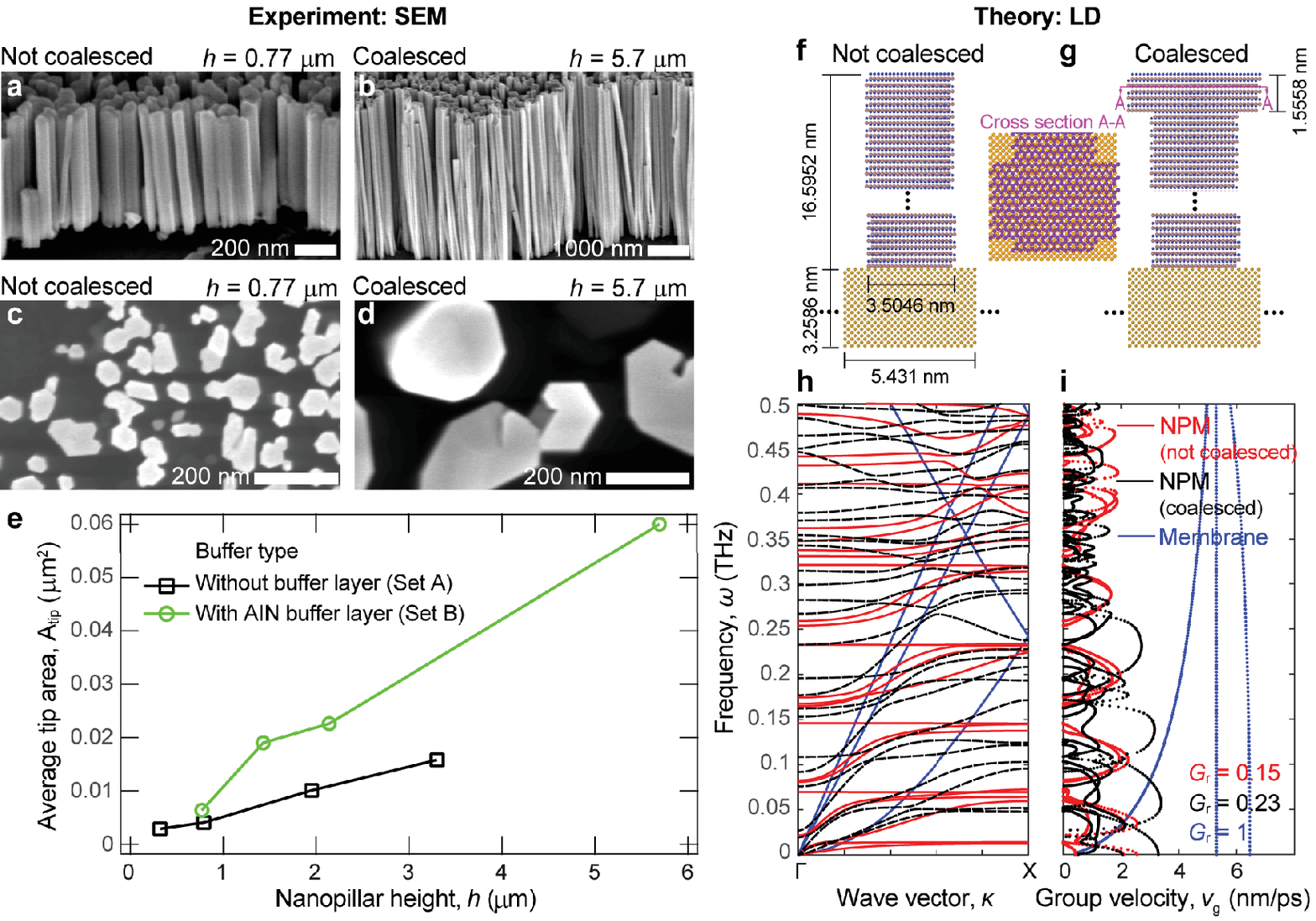}
\caption{\textbf{Nanopillar coalescence: Evidence of NPM effect.} SEM images of specimen with (a) least coalescence and (b) greatest coalescence, both in tilt view $45^\circ$.~Top views of (a) and (b) are shown in (c) and (d), respectively.~(e) Plot of average tip area versus nanopillar height showing that coalescence increased with nanopillar height. Atomic model of unit cell (f) without coalescence and (g) with coalescence (base membrane brown, nanopillar purple), and corresponding (h) phonon band structure and group-velocity distribution.~The average group velocity for NPM normalized with respect to corresponding bare membrane is shown to increase by 53~\% with coalescence.}
\label{fig:Fig4}
\end{figure*}
\indent Following the development given in Ref.~\cite{Cai2010} for bare (unpillared) membranes, the lateral thermal transport is governed by a radial heat equation with a source heating term.~We find $\Delta T(r) = T(r) - T_{\textup{amb}} = P_{\textup{abs}}{\textup{ln}}(r/R) \beta(r)/(2\pi d_{i} k_{i})$ , where $P_{\rm{abs}}$ is the absorbed power from the heating laser with beam radius $r_0$, $r$ is the radial distance from the center of the laser spot, $T_{\textup{amb}}$ is the ambient temperature, $R$ is the radius of the membrane (to the boundary where it attaches to the silicon wafer), and $d_{i}$ and $k_{i}$ are the effective conductive thickness and thermal conductivity, respectively, where $i$ represents either a bare membrane ``Mem" or a nanopillared-covered membrane ``NPM". As described in the Appendix, the radial temperature variation is small within the probe beam diameter, and thus the measured temperature difference relative to $T_{\rm{amb}}$ can be equated to $\Delta T(0)$, for which $\rm{ln}(r/R)\beta(r)$ becomes $\rm{ln}(R/r_0) +\gamma/2$, where $\gamma$ is the Euler constant = 0.57721 to five significant digits. In this study, we are primarily interested in the effects of the added nanopillars on the surface of the Si membranes. In order to single out the nanopillar effects on the thermal conductivity, we differentiate the previous equation with respect to the absorbed laser power and take the ratio of this differential expression for the specimens with nanopillars and the specimens with bare membranes, i.e., 
\begin{equation}
 \frac{\partial \Delta T_{\rm{Mem}}/\partial P_{\textup{abs}}}{\partial \Delta T_{\textup{NPM}}/\partial P_{\textup{abs}}} = \frac{k_{\textup{NPM}}{d_{\textup {NPM}}}}{k_{\textup{Mem}}{d_{\textup {Mem}}}}.
\end{equation} 
In our measurements, the power of the 325 nm beam was varied and the slope of the temperature versus absorbed power was used to derive the relative ratio of the thermal conductivities.~We convert the relative changes in $k$ to estimates of absolute thermal conductivity by multiplying a typical thermal conductivity of 200-nm thick Si membranes, 60 W/m$\cdot$K~\cite{chavez2014, cuffe2012}, by the ratio of the inverse slope of each sample to the average value of the inverse slope for membranes without nanopillars, 0.0251 mW/K.  Although $d_{\textup{NPM}}$ is greater on average than $d_{\textup{Mem}}$ because of the presence of the nanopillars, we make the assumption that these two thicknesses are equal and cancel in Eq.~(1). This assumption tends to underestimate the thermal conductivity reduction by the NPM effect.~In the SI, we discuss how surface roughness variation and heat loss to the environment are not consequential in our experiments.  \\
\indent As can be seen in Fig.~\ref{fig:Fig3}a, the thermal conductivity for the specimens displays a significant reduction as the height of the nanopillars increases, with a maximum reduction of 21~$\% \pm{0.4~\%}$.~The source of this reduction is explained by examining the phonon band structure of the NPM unit cell.~For our models, we consider a representative unit cell with a Si base width of 85 nm and thickness of 200 nm, supporting a GaN nanopillar with a square cross-section, a width of 55 nm, and a height targeted to vary from 0.5 to 4 $\mu$m.~A corresponding atomic model was created with all dimensions $\thicksim$15 times smaller for feasible computation (see Figs.~\ref{fig:Fig2}e,f and Methods).~As shown in~Fig.~\ref{fig:Fig2}g, the nanopillars fundamentally transform the membrane band structure by adding a population of localized modes that appear as horizontal lines spanning the Brillouin zone; these represent the resonance/vibron modes that couple with the underlying membrane phonon dispersion modes throughout the spectrum (the NPM effect).~The localizations manifest physically as illustrated in the atomic motion close-up inserts in Fig.~\ref{fig:Fig2}f.~The outcome is strong reductions in the phonon group velocities $v_{\textup{g}}$ and their mode participation ratios $p_{\textup{r}}$ which quantify the extent of mode localization in the NPM unit cell; see definitions in Methods.~These two factors directly contribute to reducing the in-plane thermal conductivity~\cite{Honarvar2018}.~Equilibrium MD simulations were also conducted on the same atomic-scale NPM model, followed by application of the Green-Kubo method, producing a trend similar to the experimental trend of a reduction in $k$ with nanopillar height (see green curve in~Fig.~\ref{fig:Fig3}a and Methods).~The MD results indicate a reduction of nearly 92 $\%$, which is higher than the experimental reduction because of the smaller features sizes compared to the phonon MFP distributions of Si and GaN.~This similarity in trends shows that the NPM effect describes the data we observe experimentally.   \\
\indent Unlike strategies of introducing defects that also slow electronic carrier transport, we see no negative impact on the electrical conductivity of the specimens (Fig.~\ref{fig:Fig3}b), while both sets display similar reductions in $k$.~As Figs.~\ref{fig:Fig3}b and~\ref{fig:Fig3}c show, the $\sigma$ and $S$ values for Set B are unaffected by the presence of increasingly taller nanopillars, while $k$ is reduced for all but the severely coalesced specimen.The low value of $\sigma$, around 200 S/m, is due to the low doping in the samples (see Methods and Appendix).~This data rules out the possibility of scattering-induced reductions in carrier mobility or density from the presence of the nanopillar forest. As explained previously, Set A shows an increase in electrical conductivity that we attribute to coincidental Ga diffusion and not to improvement in mobility.~The Seebeck coefficients for Set A show the typical decrease as carrier concentrations increase~\cite{wagner2007, Geballe1955, Ohishi2015}.~Thus we have clearly shown that the thermal properties and electrical properties of the nanopillared membranes have been decoupled. \\
\indent Under ideal circumstances, theory predicts that having larger nanopillars attached to the Si membranes should reduce the thermal conductivity by increasing the number of vibrons available for coupling with the base-membrane phonons.~As can be seen in Fig.~\ref{fig:Fig3}a, the initial decreases in $k$ with increasing nanopillar height reverse themselves for Set B, with the NPM effect extinguished at a nanopillar height of 5.7 ${\mu}$m.~This reversal is explained by an unavoidable coalescence of neighboring nanopillars as the nanopillar height increases.~We observe that the coalescence occurs predominantly near the tips rather than at the roots.~A comparison of two extreme cases is given in Figs.~\ref{fig:Fig4}a-d.~We quantify the coalescence by calculating an average tip area using standard image analysis techniques; the complete image set is available in the Appendix.~The tip areas plotted in Fig.~\ref{fig:Fig4}e show that most of the specimens in this study display some degree of coalescence, and the effect is significantly ($\sim3\times$) stronger for the tallest nanopillars in Set B.~The nullification of the observed NPM thermal conductivity reduction by coalescence is also seen in quasiharmonic lattice dynamics calculations, as shown in Figs.~\ref{fig:Fig4}f-i.~The phonon band structure shows that vibron states (horizontal black lines) move to higher frequencies when the nanopillars touch at the tips and thus reduce the NPM effect at the lower frequency regime which is dominant in the thermal transport~\cite{DavisHussein2014}.~Furthermore, an increase in the average group velocities across the spectrum is observed due to having less isolated nanoresonators.~These changes cause an increase in $k$ relative to nanopillars with unconnected tips, which provides further proof that the thermal conductivity reduction is due to the NPM effect and not scattering of phonons from the nanopillar roots.~More broadly, the results offer an experimental demonstration of the role of wave effects in thermal transport in nanostructures with feature sizes on the order of a few hundred nanometers, at room temperature.~This finding establishes a unique analogy with acoustics, given that the introduction of substructures to induce intrinsic local resonances has been widely utilized in the form of acoustic metamaterials~\cite{Liu2000}; here the concept is experimentally realized$-$for the first time$-$at the nanoscale for influencing the thermal conductivity.  \\ 
\section{Conclusions}
\indent The ultimate target of decoupling TE properties is to enable a route for increasing $ZT$.~In Fig.~\ref{fig:Fig3}d, we see that the NPM effect has increased the relative $ZT$ by a factor of 2.7, raising the absolute value from $0.42\times10^{-3}$ for the bare membrane to $1.12\times10^{-3}$.~The theory predicts that significantly larger enhancements are possible in more ideal specimens with larger ratio of nanopillar-to-membrane volume~\cite{Honarvar2018,Hussein_2018}.~Our results demonstrate that these gains are obtained by the NPM effect without degradation in the electrical properties of membranes.~By increasing doping in the base membrane, the numerator in the $ZT$ expression will also increase to provide significant additional gains in the $ZT$ absolute value.~Furthermore, these results have been demonstrated in base membranes with robust dimensions and in a material that is technologically advanced and inexpensive.~The enhancement through nanostructure-induced resonances would apply to other semiconductors as well, including common TE materials~\cite{Yang2020}, provided the phonon MFP distribution has significant overlap with the nanostructure features.~Together these results point to a long-sought solution to the problem of maximizing TE material performance by breaking the coupling between the thermal and electrical properties.
  
\section{Methods Section}
\subsection*{Nanopillar synthesis}
\textit{MBE growth}: GaN nanopillars were grown by catalyst-free MBE with a plasma-assisted nitrogen source onto the Si(100) device layer prior to membrane etching and release.~The SOI substrates (SEH America\footnote{Vendor is identified to adequately specify the source material.~This identification does not imply recommendation or endorsement by the National Institute of Standards and Technology, nor does it imply that the product identified is necessarily the best available for the purpose.}) had device, buried oxide, and carrier layer thicknesses of 200 nm, 380 nm, and 675 $\mu$m, respectively.~The device layer was lightly boron doped with a resistivity of 28 $\Omega$-cm; as noted above, these thin device layers are not currently available in any other doping types or concentrations. The nanopillars initially cover the entire surface of the substrate but were selectively removed with photolithography for the electrical test structures~\cite{weber2017}. Nanopillar height was varied by adjusting the nanopillar growth period, with the longest growth period being 12 h.~The ratio of the N equivalent growth rate to the Ga equivalent growth rate during nanopillar growth was 6:1 for the Set A and 3:1 for Set B.~The nanopillars were grown at approximately 810 $^{\circ}$C.~More details are provided in the Appendix. 

\subsection*{Sample fabrication}
After nanopillar growth, each 2 cm $\times$ 2 cm chip was fabricated into a testing platform to measure its thermoelectric properties.~Each completed chip yields 2 four-point electrical resistivity devices, 2 Seebeck coefficient devices, and 92 thermal conductivity test membranes ranging in nominal size from 400 $\mu$m $\times$ 400 $\mu$m to 700 $\mu$m $\times$ 700 $\mu$m.~Ohmic contact pads were formed using 20 nm Ti/200 nm Al metal stacks annealed in argon at 500 $^{\circ}$C for 1 minute.  

\subsection*{Thermoelectric metrology}
\textit{Raman thermometry:}~We used a 325-nm He-Cd laser as a heating source that was propagating anti-parallel to a low intensity 532-nm laser used as a Raman probe.~The nanopillared Si membranes were positioned such that the side with nanopillars was exposed to the low intensity 532-nm Raman probe, while the 325-nm beam was absorbed on the unpatterned side of the membrane.~This optical alignment allowed for more accurate estimation of absorbed laser power due to the $\sim$60-nm absorption depth at the 325-nm wavelength, precluding transmission to the nanopillars on the opposite side of the membrane.~The beam diameters at the $1/e^{2}$ points were 25 $\mu$m and 0.8 $\mu$m for the 325-nm and 532-nm lasers, respectively. The nanopillars are transparent to the green probe beam though some scattering occurred as the beam passed through them.~The reflectance $R$ of the bottom side of the membranes was measured to be 0.57 for the UV beam, and the absorbed beam power was calculated as the incident beam power multiplied by $(1-R)$.~The beam power was measured with an optical power meter close to where it impinged on the specimen and then corrected for transmission of the intervening optics.~The temperature dependence of the Si Raman peak was calibrated by heating a Si chip with a strip heater and measuring its temperature with a thermocouple while acquiring Raman data. The resulting data was fit with the quadratic equation $T(^\circ \rm{C}) = 23.2-50.4(\rm{\Delta}\nu-1.1 (\Delta\nu)^{2}$) where $\Delta\nu$ is the temperature-induced shift in the Raman peak position in wavenumbers (cm$^{-1}$). The linear term of this equation agrees well with previous evaluations that report $\Delta T/\Delta\nu =-46$~${\rm K}/{\rm cm}^{-1}$, initially by the work of Mendez and Cardona and verified by others including Reparaz et al.~\cite{reparaz2014,cardona1984}. \\

\noindent \textit{Seebeck coefficient:}~The Seebeck coefficient measurement was performed via a steady-state method with the geometry shown in the Appendix, Fig.~\ref{fig:FigS4}.~Two meandering Ti/Al wires were lithographically defined 10 $\mu$m from the Si device layer to serve as thermometers with a $\sim$100-$\Omega$ resistance.~Prior to measurement, both resistors $R_1$ and $R_2$ were calibrated to within 0.1 K.~An additional pair of Ti/Al wires was patterned in direct contact with either end of the Si device layer to measure the Seebeck voltage.~Two 1-k$\Omega$ chip resistors, serving as heaters, were glued to one end of the chip and used to provide a thermal gradient along the length of Si device layer.~The heaters provided up to 25 mW of power yielding a maximum $\Delta T$ of ~3.5 K.~The heater current, thermopower voltage $V_{\rm th}$, temperatures at $R_1$ and $R_2$, and temperature gradient across the Si device layer $\Delta T$ were recorded as a function of time with initial sample temperature at 277 K.~All calibrations and measurements were performed in ice water to maintain a constant bath temperature. \\

\noindent \textit{Electrical resistivity:}~The electrical resistivity was measured using a standard four-point probe test structure shown in the Appendix, Fig.~\ref{fig:FigS4}.~The quantity $\Delta$V across the two inner contacts was measured as a function of current across the two outer contacts over the range of 0 nA$-$100 nA.~All tested devices showed a linear, ohmic response, allowing for resistivity $\rho$ to be calculated from the membrane  width $w$, thickness $t$, length $L$, and measured resistance $R$ as $\rho=Rwt/L$.

\subsection*{Atomic models}
The theoretical investigations are based on atomic models comprising a Si membrane with GaN nanopillars standing on the surface. Both material portions were modeled as single crystals under room-temperature equilibrium conditions.~The Tersoff potential was used for the interatomic interactions.~The parameters of the Si-Si and Ga-N interactions were taken from Refs.~\cite{Tersoff1988} and~\cite{Nord2003}, respectively.~For the Si-Ga and Si-N interactions, the potential parameters were mixed following the Tersoff multicomponent combination rules~\cite{Tersoff1989}.~Two sizes of NPMs were investigated: one that is nearly 15 times smaller than a nominal experimental unit cell (shown in Fig.~\ref{fig:Fig2}f, right), and a smaller version for the coalescence investigation (shown in Fig.~\ref{fig:Fig4}g).~In the model of the coalesced NPM, the top of the nanopillar was laterally extended to partially connect with adjacent nanopillars.~This was done by adding three primitive-cell layers of GaN around the tip of the nanopillar forming a cross-like cross section when viewed from the top (cut view A-A in Fig.~\ref{fig:Fig4}g). \\

\noindent \textit{Lattice dynamics calculations:}~The phonon band structures for the examined GaN-on-Si NPM unit cells were obtained by solving the quasiharmonic lattice dynamics eigenvalue problem using the GULP software~\cite{Gale2003}.~Bloch periodic boundary conditions were applied along in-plane directions and free boundary conditions were applied in the $z$ direction and around the nanopillar.~The phonon frequencies were computed at a set of allowed wave vectors ranging from $\Gamma$ to $X$ in the Brillouin zone with a resolution of 128 points.~There are $3N$ phonon branches in the band structure, where $N$ is the total number of atoms in the unit cell. \\
\indent The average group velocity ratio ${G}_{\textup{r}}$ is a quantity that characterizes the reduction in the group velocities across the entire phonon spectrum~\cite{Honarvar2018}.~It is defined as ${G}_{\textup{r}}$ = ${G}_{\textup{NPM}}$ /${G}_{\textup{Mem}}$, where $G_i$ is the average group velocity of either an NPM or a membrane calculated by ${G_i} = (1/{n_\kappa n_m})\sum_{\kappa}^{n_\kappa} \sum_{m}^{n_m} v_{\textup{g}}(\kappa, m)$.~Here, $\kappa$ is the wave number (scalar component of the wave vector $\boldsymbol{\kappa}$ along the $\Gamma$$-$$X$ direction), $m$ is the branch number, $n_\kappa$ is the number of wave-number points considered, and $n_m=3N$ is the total number of phonon branches.~The  group velocity $v_{\textup{g}}(\kappa, m)$ is defined as the slope of the phonon frequency with respect to the wave number $\kappa$ for branch $m$.\\
\indent ~For characterization of nanopillar resonant mode localization, we examine the mode shape corresponding to each point in the phonon band structure. We then compute the mode participation ratio $p_{\rm r}$, which is defined for a mode at wave vector $\bm{\kappa}$ and branch number $m$ by Ref.~\cite{Wei2015,Honarvar2018}
\begin{equation}
\label{PR}
{{p_{\rm r}}(\bm{\kappa}, m)}  =  
\dfrac{1}{N \sum_{i=1}^{N}[\sum_{j=1}^{3}\boldsymbol{\phi}^*_{i j} ( \bm{\kappa},m)\boldsymbol{\phi}_{i j} ( \bm{\kappa},m)]^{2}},
\end{equation}
\noindent where $\boldsymbol{\phi}_{i j} ( \bm{\kappa},m)$ is the displacement component corresponding to atom $i$ and direction $j$ of the normalized mode shape.~The formula comprises two summations.~The first is over the total number of atoms $N$ in a unit cell, i.e., $N = N_{\rm Base} + N_{\rm Pillar}$ for an NPM, where $N_{\rm Base}$ is the number of atoms in the base membrane and $N_{\rm Pillar}$ is the number of atoms in the nanopillar.~The second summation is over the three directions of motion per atom.~The inverse of this quantity $p_{\rm r}$ indicates the degree of modal localization over the entire unit cell considered without being specific to a particular region, e.g., the nanopillar or base membrane portion of an NPM unit cell.~This calculation is performed for both an NPM and a bare membrane.~In an NPM, a large number of the modes exhibit high concentrations of vibrations in the nanopillar portion, yielding a low value of $p_{\rm r}$.\\

\noindent \textit{Molecular dynamics simulations:}~Equilibrium molecular dynamics (EMD) simulations were executed to predict the in-plane lattice thermal conductivity of the GaN-on-Si NPMs sized at nearly 1/15 of the nominal experimental unit cell, with the height of the nanopillar being varied (see atomic model dimensions in Fig.~\ref{fig:Fig2}).~A single unit cell was used as the simulation cell with periodic boundary conditions applied along the $x$ and $y$ directions and free boundary condition applied in $z$ direction and around the nanopillar.~The empirical interatomic potentials were identical to those used in the LD calculations.~The time integration step was set as $\Delta t = 0.5$~fs. First, a canonical ensemble MD with a Langevin heat reservoir was allowed to run for 0.3~ns to enable the whole system to reach equilibrium at 300~K.~Then, a microcanonical ensemble (NVE) was run for 3~ns; meanwhile, the heat current was recorded at each time step. At the end of the simulations, the thermal conductivity was calculated by the Green-Kubo formula,~\cite{schelling2002comparison} $k ={1}/{(2Vk_{B} T^{2})} \int_0^{\infty} \langle \bm{J}(\tau)\centerdot \bm{J}(\tau) \rangle {\rm d} \tau$ where $k_{B}$ is the Boltzmann constant, $V$ is the system material volume, and $\bm{J}$ is the heat flux along the direction of transport.~Finally, the thermal conductivity was averaged over the two in-plane directions over six simulations with different initial velocities (i.e., a total of 12 cases), and the statistical errors were obtained according to the method described in~\cite{schelling2002comparison}. All EMD simulations were performed in LAMMPS~\cite{plimpton1995fast}.

\section*{Acknowledgments}
This research was partially supported by the Advanced Research Projects Agency$-$Energy (ARPA-E) under grant number DE-AR0001056.\\

\appendix*
\section*{APPENDIX}

\renewcommand{\thefigure}{A\arabic{figure}}
\renewcommand{\thetable}{A\arabic{table}}
\renewcommand{\theequation}{A\arabic{equation}}

\setcounter{figure}{0}    
\setcounter{table}{0}    
\setcounter{equation}{0} 

This supplemental information document contains tables of data from the figures used in the main article, further details of the methods and results, including a full set of SEM images of the specimens used in the study and examples of the Raman temperature vs. absorbed power curves, discussion of possible additional heat transport mechanisms, and discussion of the Al and Ga diffusion into the membrane.~Commercial equipment and instruments are identified in order to adequately specify certain procedures.  In no case does such identification imply recommendation or endorsement by the National Institute of Standards and Technology, nor does it imply that the products identified are necessarily the best available for the purpose. 

\section{Tabular data}
This section contains details of specimen synthesis and morphology and data from Fig.~3 in the main article.

\begin{table}[ht]
\begin{center}
  \caption{Specimen characteristics for Sets A and B.}
  \label{tab:table1}
  \begin{tabular}{l|c|c|c}
    Run No. &  Nanopillar &Growth&$k_{\rm NPM}/k_{\rm Mem}$ \\
     &  height ($\mu$m) & time (h) & \\
    \hline
    \hline
    Bare (A) & NA & NA & 0.99 $\pm$ 0.06 \\
    D420 (A) & 0.324 & 1.7 & 1.03 $\pm$ 0.05 \\
    D421 (A) & 0.785  & 3.2 & 0.98 $\pm$ 0.04 \\
    D422 (A) & 1.950 & 6.2 & 0.85 $\pm$ 0.02 \\
    D423 (A) & 3.300 & 12.2 & 0.79 $\pm$ 0.01 \\
    \hline
    D442 (B) & NA & 0.083 & 1.01 $\pm$ 0.01 \\
    D469 (B) & 0.77 & 2.6 & 0.93 $\pm$ 0.04  \\
    D480 (B) & 1.430 & 2.6 & 0.80 $\pm$ 0.02  \\
    D443 (B) & 2.140 & 5.1 & 0.87 $\pm$ 0.01 \\
    D481 (B) & 5.700 & 10.1 & 0.98 $\pm$ 0.01 \\
  \end{tabular}
\end{center}
\end{table}

\begin{table}[ht]
\begin{center}
  \caption{Nanopillar dimensions for Sets A and B.}
  \label{tab:table2}
  \begin{tabular}{l|c|c|c|c}
    Run No. &  Nanopillar &Root diam.& Fill & Avg. tip  \\
     &  height ($\mu$m) & (nm) & fraction & area ($\mu$m$^2$)\\
    \hline
    \hline
    D420 (A) & 0.324 $\pm$ 0.035 & 30 $\pm$ 4 &  0.33 & 0.0029 \\
    D421 (A) & 0.785 $\pm$ 0.130 & 40 $\pm$ 11 &  0.25 & 0.0041 \\
    D422 (A) & 1.950 $\pm$ 0.200 & 50 $\pm$ 11 &  0.38 & 0.010 \\
    D423 (A) & 3.300 $\pm$ 0.320 & 75 $\pm$ 22 &  0.35 & 0.016 \\
    \hline
    D469 (B) & 0.77 $\pm$ 0.050 & 65 $\pm$ 14 & 0.39 & 0.0064 \\
    D480 (B) & 1.430 $\pm$ 0.080 & 70 $\pm$ 25 &  0.52 & 0.019 \\
    D443 (B) & 2.140 $\pm$ 0.040 & 60 $\pm$ 12 &  0.54 & 0.023 \\
    D481 (B) & 5.700 $\pm$ 0.140 & 130 $\pm$ 36 &  0.39 & 0.060 \\
  \end{tabular}
\end{center}
\end{table}

\begin{table}[ht]
\begin{center}
  \caption{Thermal conductivity, electrical conductivity, and Seebeck coefficient data from Fig.~3 of the main article.~The thermal conductivity values were obtained by multiplying the relative values in Table~\ref{tab:table1} by $60~{\rm W/m\cdot K}$, as discussed in the text. }
  \label{tab:table3}
  \begin{tabular}{l|c|c|c|c}
    Run No. &  Nanopillar & Thermal cond. & Electrical cond. & Seebeck coef.\\
     &  height ($\mu$m) & $k_{\rm NPM} (\rm W /m\cdot K)$ & $\sigma$ (S/m) & ($\mu$V/K)\\
    \hline
    \hline
    Bare (A) & NA & 59.6 $\pm$ 3.4 & 3.6 $\pm$ 0.5 & 1500 $\pm$ 12  \\
    D420 (A) & 0.324 & 61.9 $\pm$ 2.8 &  2.8 $\pm$ 0.2 & 955 $\pm$ 47 \\
    D421 (A) & 0.785 & 58.5 $\pm$ 2.6 &  41.4 $\pm$ 5 & 800 $\pm$ 8 \\
    D422 (A) & 1.950 & 50.9 $\pm$ 1.0 &  103 $\pm$ 30 & 730 $\pm$ 5 \\
    D423 (A) & 3.300 & 47.2 $\pm$ 0.5 &  233 $\pm$ 30 & 610 $\pm$ 1.6 \\
    \hline
    D442 (B) & NA & 60.4 $\pm$ 0.6 &  170 $\pm$ 20 & 711 $\pm$ 1 \\
    D469 (B) & 0.77 & 55.6 $\pm$ 2.2 &  no data & no data  \\
    D480 (B) & 1.430 & 47.7 $\pm$ 1.0 &  150 $\pm$ 30 & 676 $\pm$ 2  \\
    D443 (B) & 2.140 & 52.4 $\pm$ 0.7 &  320 $\pm$ 50 & 788 $\pm$ 1  \\
    D481 (B) & 5.700 & 58.8 $\pm$ 0.8 &  220 $\pm$ 50 & 719 $\pm$ 3  \\
  \end{tabular}
\end{center}
\end{table}

\section{Nanopillar synthesis}
SOI substrates (SEH America) were diced into 2 cm by 2 cm squares prior to growth and processing.~Immediately prior to loading for growth, the substrates were cleaned with solvents, oxygen plasma in a reactive ion etching system, and approximately two-minute exposure to hydrogen fluoride (HF) vapor.  The substrates were outgassed three times at successively higher temperatures in the MBE loadlock chamber, preparation chamber, and growth chamber, with the final outgas reaching between 870 $^{\circ}$C and 890 $^{\circ}$C.~The Si surface initially displayed a 1$\times$1 reflection high energy electron diffraction (RHEED) pattern that brightened and sharpened as surface oxides desorbed.~Around 720 $^{\circ}$C, the 1$\times$1 pattern changed into a 2$\times$1 pattern.~Substrate temperatures throughout the growth were measured with an estimated uncertainty of 8 $^{\circ}$C using a back-side pyrometer described elsewhere~\cite{bertness2014}. Element fluxes were estimated from separate growth calibration runs in which the growth rate of planar films was derived from optical interference fringes at a variety of different Ga and Al beam equivalent pressures under group-III-limited conditions.~The nitrogen-limited growth rate for a variety of plasma conditions was estimated from the transition from a spotty to streaky RHEED pattern as the gallium flux was increased~\cite{brubaker2016}.~For specimen Set A (no buffer layer), the Ga and N equivalent planar growth rates were 110 $\pm$ 10 nm/h and 650 $\pm$ 30 nm/h, respectively. For these runs, the Si surface was exposed to the N plasma for 60 s at 740 $^{\circ}$C, then nanopillar growth was initiated at a low temperature of 700 $^{\circ}$C for 12 minutes. The remainder of the nanopillar growth took place at 810 $^{\circ}$C.  For specimen Set B (AlN buffer layer), the Ga and N equivalent planar growth rates were 250 $\pm$ 20 nm/h and 650 $\pm$ 30 nm/h, respectively.  Buffer layer deposition was preceded by a 15~s N plasma exposure and a 1~s Al exposure.  The buffer layer itself was deposited at an actual growth rate of 165 $\pm$ 20 nm/h with the N flux lower than the Al flux to promote formation of a N-polar surface.~The growth temperature for the buffer layer was 850 $^{\circ}$C, where we estimate that approximately half of the Al reevaporated rather than be incorporated.~GaN nanopillar growth was initiated on the AlN buffer layers at 770 $^{\circ}$C for 5 minutes before increasing to 810 $^{\circ}$C for the main nanopillar growth stage.  The GaN nanopillars grown without a buffer layer were not intentionally doped, leading to an n-type carrier concentration below $1 \times 10^{16}$ cm$^{-3}$.  The GaN nanopillars on AlN were lightly doped with Si, and thus had an n-type carrier concentration in the 10$^{17}$ cm$^{-3}$ range.\\

\begin{figure} [!h]
\begin{center}
\includegraphics[scale=0.65]{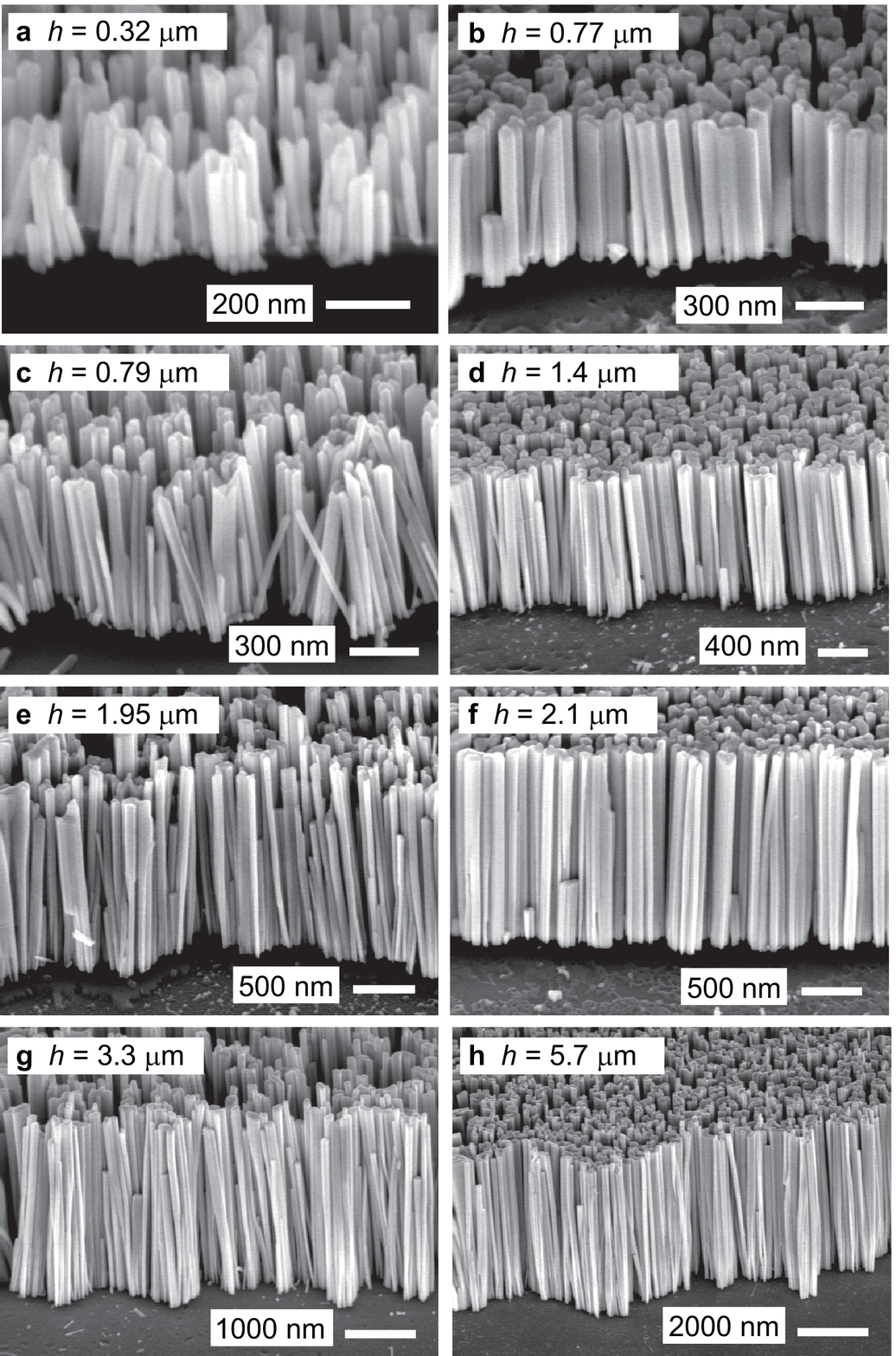}
\caption{SEM images of specimens in this study taken from a tilt angle of $45 ^{\circ}$. Images in the left column portray the specimen set grown without a buffer layer (Set A), and the right column portrays the specimen set grown with an AlN buffer layer (Set B). Run numbers are (a) D420, (b) D469, (c) D421, (d) D480, (e) D422, (f) D443, (g) D423, (h) D481.}
\label{fig:FigS1}
\end{center}
\end{figure}
\begin{figure}[!h]
\begin{center}
\includegraphics[scale=0.65]{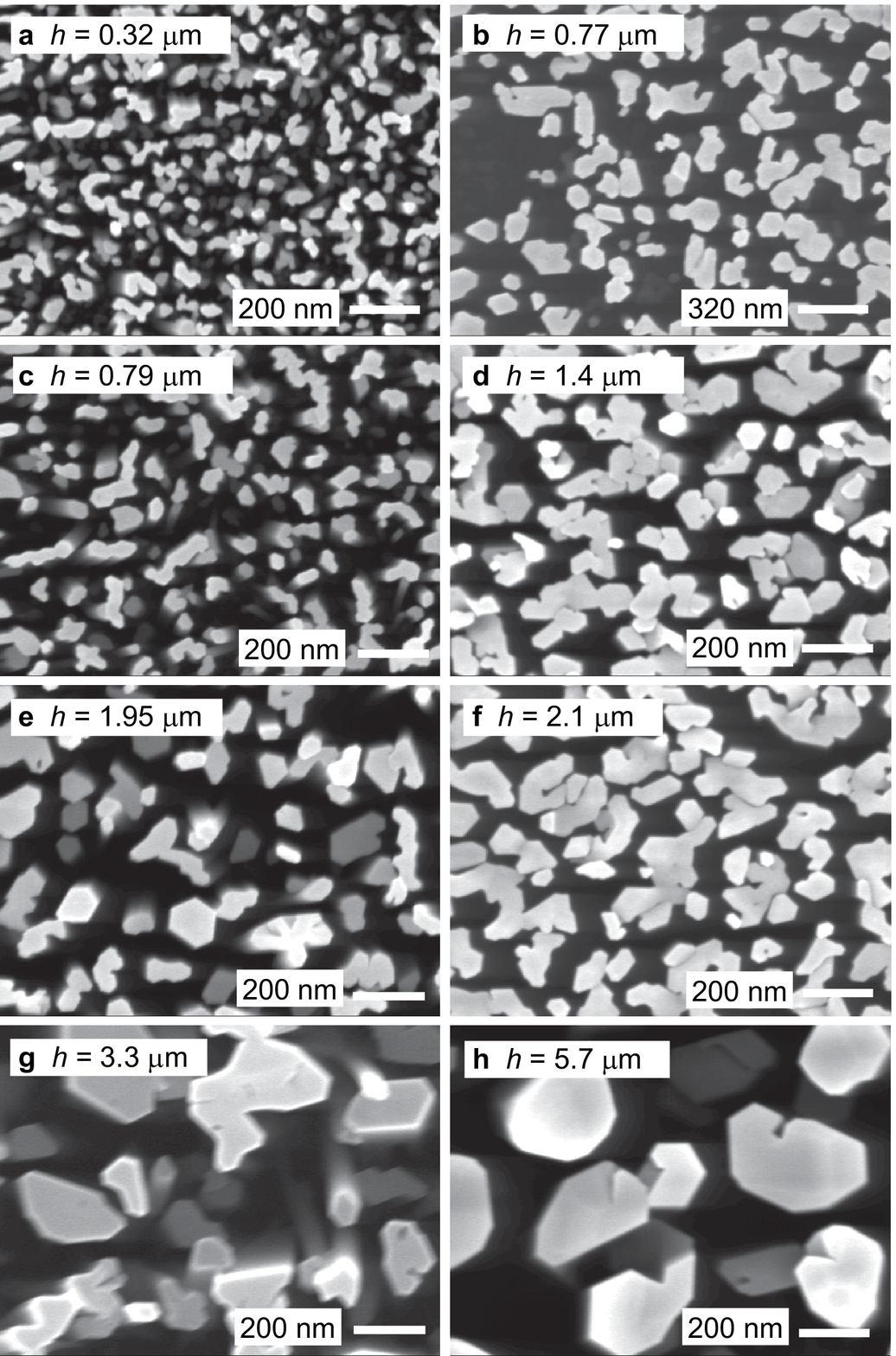}
\caption{SEM images of specimens in this study viewed from the top. Images in the left column portray the specimen Set A, and the right column portrays the specimen Set B. Run numbers are (a) D420, (b) D469, (c) D421, (d) D480, (e) D422, (f) D443, (g) D423, (h) D481.}
\label{fig:FigS2}
\end{center}
\end{figure}
\begin{figure}[!h]
\begin{center}
\includegraphics[scale=0.9]{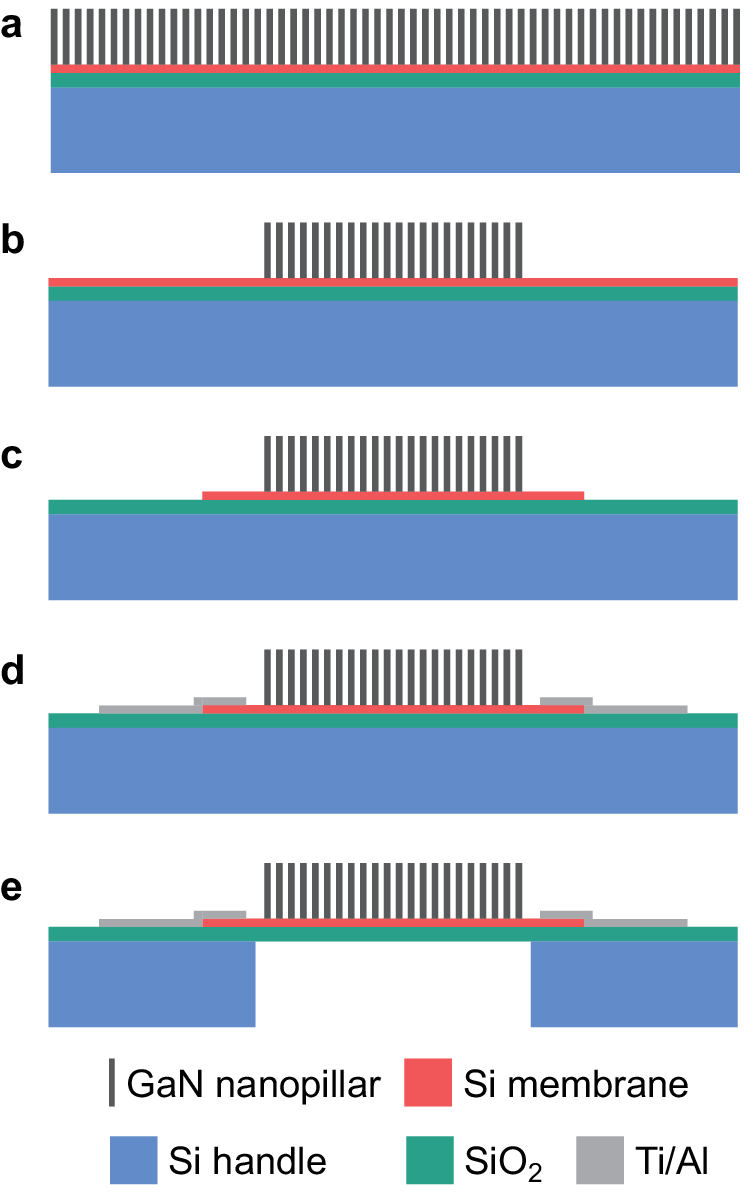}
\caption{Fabrication flow diagram.}
\label{fig:FigS3}
\end{center}
\end{figure}
\begin{figure} [!h]
\begin{center}
\includegraphics[scale=0.8]{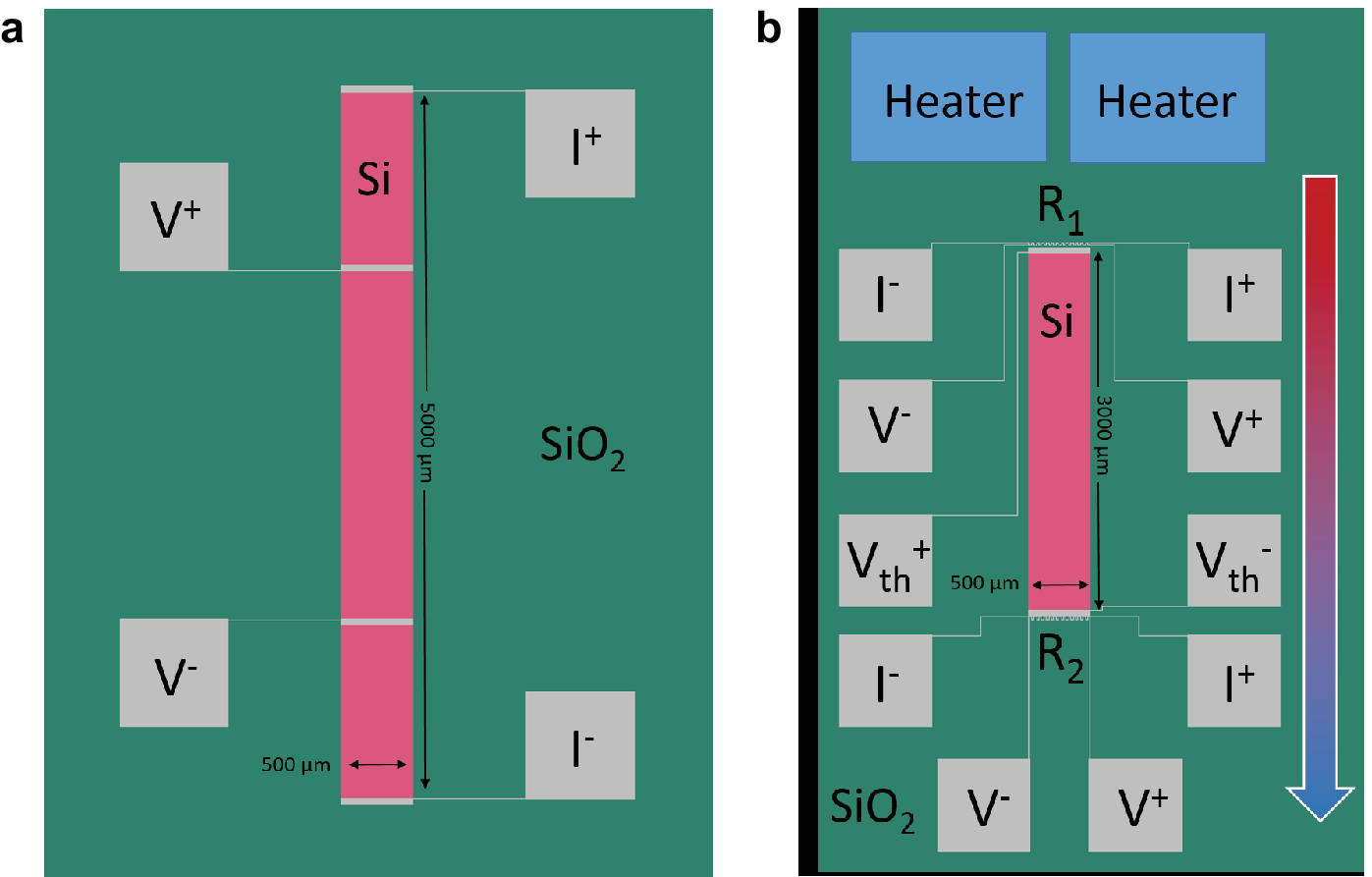}
\caption{Electrical patterns for measuring (a) electrical conductivity, and (b) Seebeck coefficient.}
\label{fig:FigS4}
\end{center}
\end{figure}
\begin{figure}[!h]
\begin{center}
\includegraphics[scale=1]{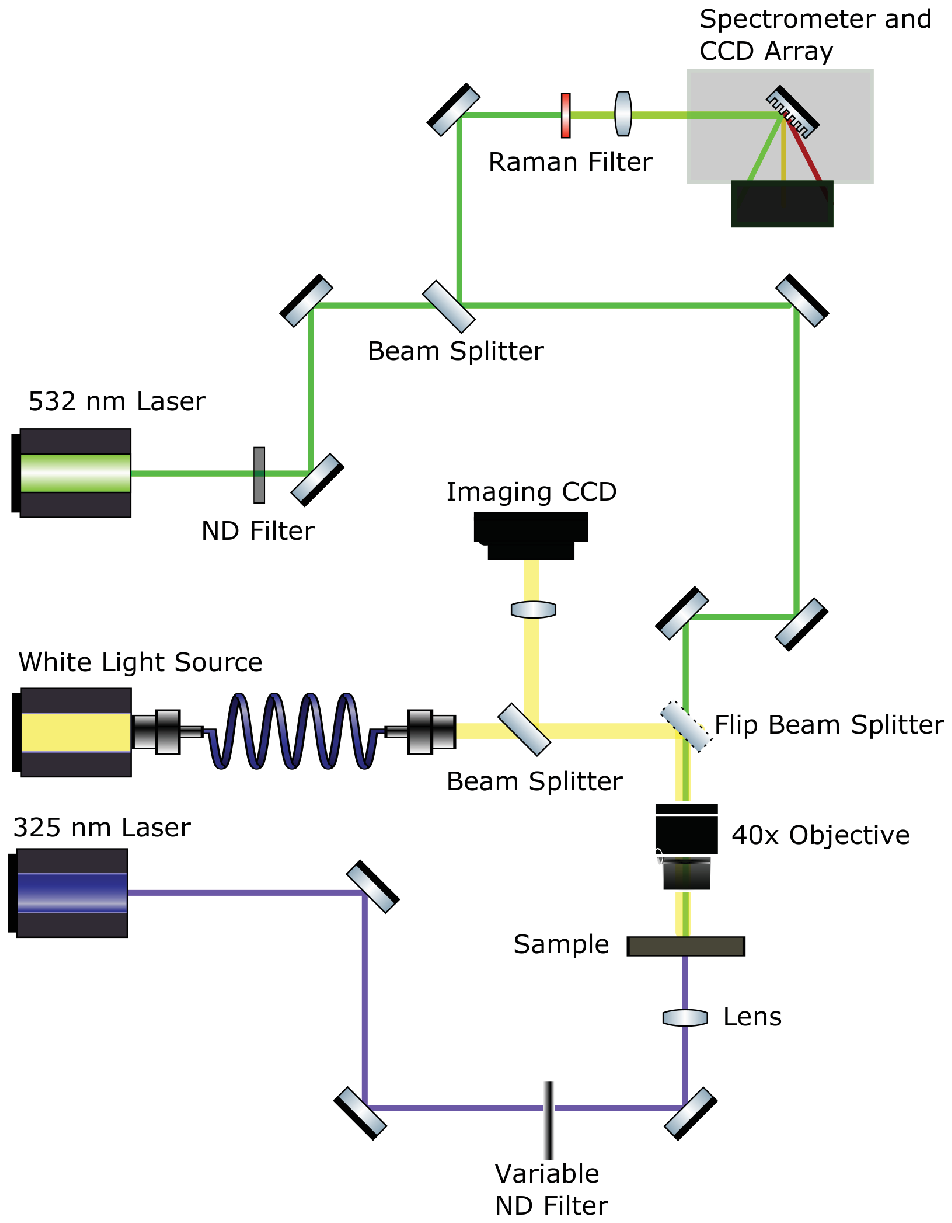}
\caption{Raman temperature measurement schematic.}
\label{fig:FigS5}
\end{center}
\end{figure}
As shown in the main article, the nanopillars tended to get thicker along their length and coalesce near the tips, and this coalescence increased with nanopillar height.  The average tip area is an approximate indicator of the degree of coalescence.  For the specimen Set B (with AlN buffer layers), the coalescence occurred faster as a function of height.  This change is most likely due to the higher Ga flux during nanopillar growth for this series. The flux was increased to increase the fill fraction of the nanopillars, defined as the ratio of the total tip area in the top view images relative to the corresponding image area, but also increased coalescence at the tips.~As can be seen in Table~\ref{tab:table2}  and Fig.~\ref{fig:FigS1}, in general the density of the nanopillars at their roots is more similar than the tip area from run to run within a set. SEM images for the both specimen sets are given in Figs.~\ref{fig:FigS1} and ~\ref{fig:FigS2}.

\section{Specimen fabrication}
The fabrication steps are shown in Fig.~\ref{fig:FigS3}.~Photolithography is used to protect the nanopillar forest with a 7 $\mu$m thick resist coating while exposed nanopillars are removed via sonication in de-ionized water.  Next, the Si device layer is patterned and etched down to the buried oxide layer with a 5-cycle deep reactive-ion etch (DRIE) using SF$_6$.~The patterns for the electrical conductivity and Seebeck coefficient measurements are in Fig.~\ref{fig:FigS4}a and~\ref{fig:FigS4}b, respectively.~A 30 s CF$_4$ dry etch is used to remove any native oxide on the Si before a 20 nm Ti/ 200 nm Al metal stack is deposited to form probe contacts and thermometers.  Deposition is followed by a 1 min 500 $^\circ$C anneal in argon to ensure ohmic contacts.~The top-side of the chip is then wax bonded to a sapphire carrier wafer to protect the nanopillars and Si device layer during back-side processing.~The membranes for thermal conductivity measurements are lithographically patterned and the Si handle layer etched with a 900 cycle DRIE followed by a 15 min HF vapor phase etch to remove the buried oxide.~The carrier wafer is then suspended upside down in a beaker of acetone to dissolve the wax and release the completed test devices.

\section{Raman thermometry}
A schematic of the Raman thermometry experiment is shown in Fig.~\ref{fig:FigS5}.~The setup consists of a Coherent Verdi V6 532 nm CW laser that is attenuated with neutral density (ND) filters to an average power of $\sim$ 300 $\mu$W at the sample. An infinity-corrected 40$\times$ Nikon objective with a numerical aperture of 0.60 was employed to focus the 532 nm beam on the samples with a spot diameter of $\sim$0.8 $\mu$m. The light is then collected in a backscattering geometry and sent through a Raman filter (Semrock RazorEdge Filter) and then into a 0.5 m spectrometer (Acton Spectra Pro 2500i) that disperses the Stokes-shifted Raman light with an 1800 groove/mm grating onto a liquid nitrogen cooled CCD array, providing sub cm$^{-1}$ energy resolution. In order to heat the samples for thermometry studies, we used a 325 nm He-Cd CW laser (Kimmon) and a variety of ND filter combinations to control incident laser fluence of the pump beam. The pump beam is collinear and propagating antiparallel to the 532 nm Raman probe beam.~The 325 nm laser beam is focused onto the membranes with a 2.0-cm focal length lens mounted on a micrometer stage.~Alignment of the two beams was tested by observing the Raman peak shift while making minor adjustments to the 325-nm lens position.  The spot diameter, defined at $1/e^2$ intensity locations, was measured to be  $\sim$25 $\mu$m at the focal point by passing a knife edge through the beam and fitting the transmitted intensity to an error function.~The pump beam average incident power was varied between 0 to 8.5 mW, resulting in fluences ranging from $\sim$0 to 3.5 kW/cm$^{2}$.~A flip-mounted pellicle beam splitter was used in conjunction with white light illumination source to image the sample and verify the Raman probe was proper centered on suspended membrane prior to each thermometry experiment.

\begin{figure} [!t]
\begin{center}
\includegraphics[scale=0.75]{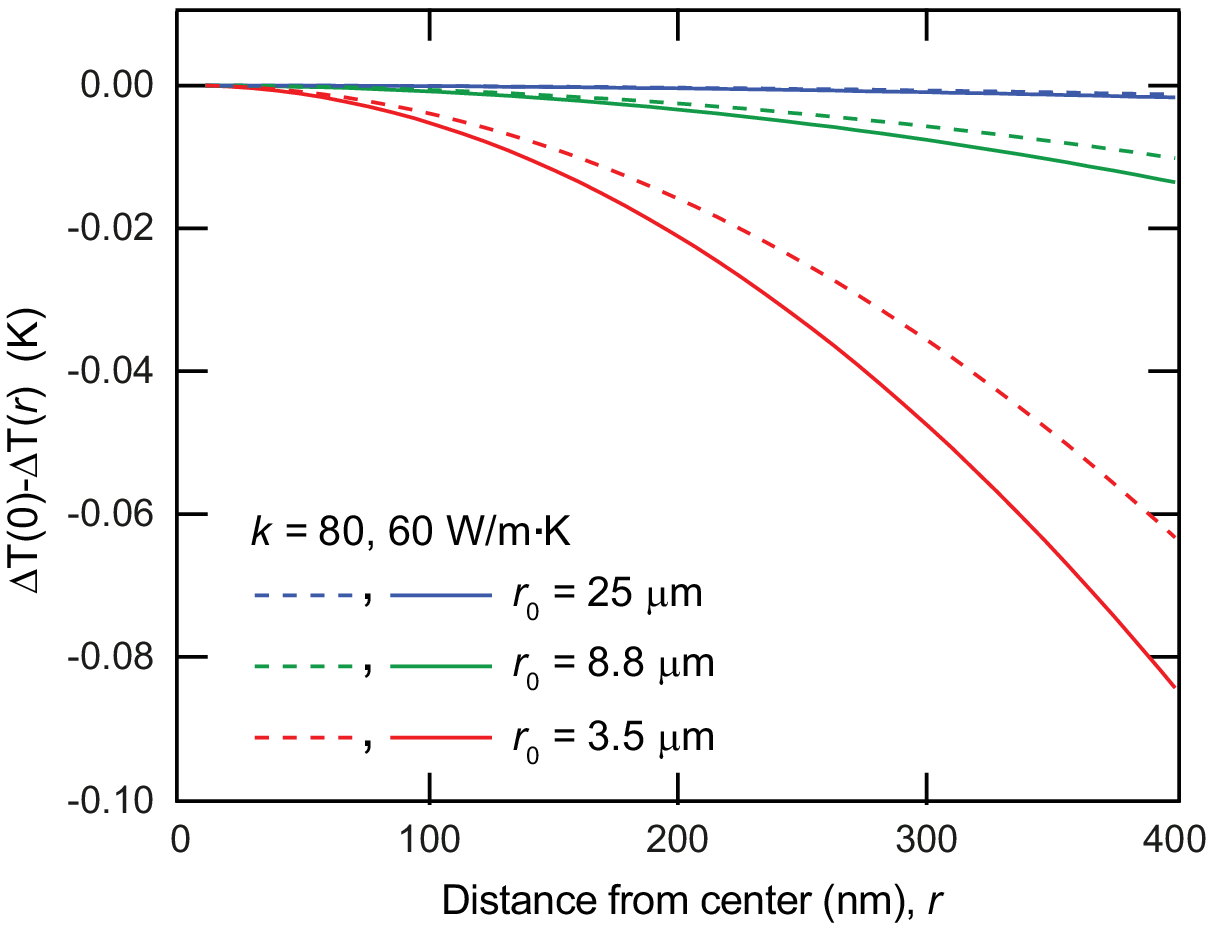}
\caption{Radial variation of $\Delta T$ relative to its value at $r = 0$ as a function of radius for representative values of membrane thermal conductivity $k$ and 325~nm beam radii $r_0$. }
\label{fig:FigS6}
\end{center}
\end{figure}

In the main text, we utilize the following radial temperature profile (see Eq.~12 in Ref.~\cite{Cai2010}): 
\begin{equation}
\label{equation:eqnS1}
\Delta T = T(r) - T_{\textup{amb}} = P_{\textup{abs}} {\textup{ln}}(r/R)  \beta(r)/(2\pi d k),
\end{equation}
where $P_{\textup{abs}}$ is the absorbed power from the 325-nm laser, $T_{\textup{amb}}$ is the ambient temperature, $R$ is the radius of the membrane, $d$ is the effective conductive thickness, and $k$ is the membrane thermal conductivity.~We evaluate some of the features of this profile to confirm our assumption that the probe beam is sampling a region with approximately uniform temperature.~We have made the assumption that the edge of the membrane is at ambient temperature for large membranes like those we use here, with $R$ on the order of 375 $\mu$m.  The function $\beta$ is defined as
\begin{equation}
\label{equation:eqnS2}
 \beta(r) =  1 +\frac{\textup{Ei}(-r^2/r_{0}^2) - \textup{Ei}(-R^2/r_{0}^2)}{2 \textup{ln} (R/r)} ,
\end{equation} 
where $R$ is the radius of the membrane, and Ei is the exponential integral function~\cite{ExpInt}. The radius $r_{0}$ of the heating laser beam is defined for an intensity distribution $ I = (P /\pi r_{0}^2 ) \textup{exp} (-r^2/r_{0}^2)$. Note that this definition of $r_{0}$ omits the conventional factor of 2 in the argument of the exponential for a Gaussian beam intensity distribution, and thus our measured beam diameter of 25~$\mu$m corresponds to $r_{0}$ = 8.8 $\mu$m. Ei$(x)$ becomes negligible for large negative arguments, and hence for $R >> r_{0}$, as is the case here, the second term in the numerator can be omitted. The behavior of Ei near $ r = 0$ can be evaluated using a Taylor series expansion for negative real arguments~\cite{Harris1957}:
\begin{equation}
\label{equation:eqnS3}
 \textup{Ei}(x) = \gamma + \textup{ln}|x| + x + \frac{x^2}{2\cdot2!} + \frac{x^3}{3\cdot3!}  ... 
\end{equation} 
Evaluating the temperature model numerically, we find that the temperature within the radius of the probe beam, 400 nm, is uniform within 0.02 K for 325-nm spot diameters ranging from 25 to 71 $\mu$m and uniform within 0.1~K for a smaller spot diameter of 10~$\mu$m (see Fig.~\ref{fig:FigS6}).~We can also evaluate the limiting behavior of the radial portion of the temperature profile near $r = 0$ by substituting the first two terms of the Taylor expansion as follows:
\begin{equation}
\label{equation:eqnS4}
\begin{split}
    \textup{ln}(R/r) \beta (r) &= \textup{ln}(R/r) \left(1 + \frac{\gamma + \textup{ln}|-r^2/r_0^2|}{2\textup{ln}(R/r) }\right) \\
     &= \textup{ln}(R/r) \left( \frac{2\textup{ln}(R/r) +\gamma +2\textup{ln}(r/r_0)}{2\textup{ln}(R/r)}\right) \\
     &= \gamma/2 + \textup{ln}(R/r) + \textup{ln}(r/r_0)\\
     &= \gamma/2 + \textup{ln}(R/r_0)
\end{split}
\end{equation} 

This temperature model can also be used to estimate the inverse slope of $\Delta T$ versus $P_{\textup{abs}}$ evaluated at the center of the 325-nm beam, which we measured to be 0.025~mW/K in bare membranes.~As shown in Table~\ref{tab:table4}, the measured value is reproduced by a combination of a membrane thermal conductivity of 60~W/m$\cdot$K and a beam diameter of 71 $\mu$m, or a membrane thermal conductivity of 80~W/m$\cdot$K and a beam diameter of 25 $\mu$m.~The difficulty in measuring the beam diameter accurately and aligning the two beams perfectly translates into difficulty in determining the absolute thermal conductivity of the membrane; hence we primarily use the Raman thermometry method for determining relative changes in the membrane thermal conductivity. 

\begin{table}[ht]
\begin{center}
  \caption{Temperature model outputs for representative membrane and beam parameters. $\Delta T$ is evaluated at $r = 0$ and $P_{\textup{abs}}$ = 1~mW.}
  \label{tab:table4}
  \begin{tabular}{l|c|c|c|c|c}
    $r_{0}$ &  Beam & \multicolumn{2}{c|}{$k$} & \multicolumn{2}{c}{$k$}   \\
    ($\mu$m) &  diam. & \multicolumn{2}{c|}{60 W/m$\cdot$K}  & \multicolumn{2}{c}{80 W/m$\cdot$K}\\
   & ($\mu$m) & Inv. Slope & $\Delta T$ (K) & Inv. Slope&$\Delta T$ (K)\\ 
   \hline
    \hline
    3.54 & 10 & 0.015 &  65.7 & 0.020 & 49.3 \\
    8.8 & 25 & 0.019 &  53.5 & 0.025 & 40.1 \\
    25 & 71 & 0.025 &  39.7 & 0.034 & 29.8 \\
  \end{tabular}
\end{center}
\end{table}

\begin{figure} [!t]
\begin{center}
\includegraphics[scale=0.75]{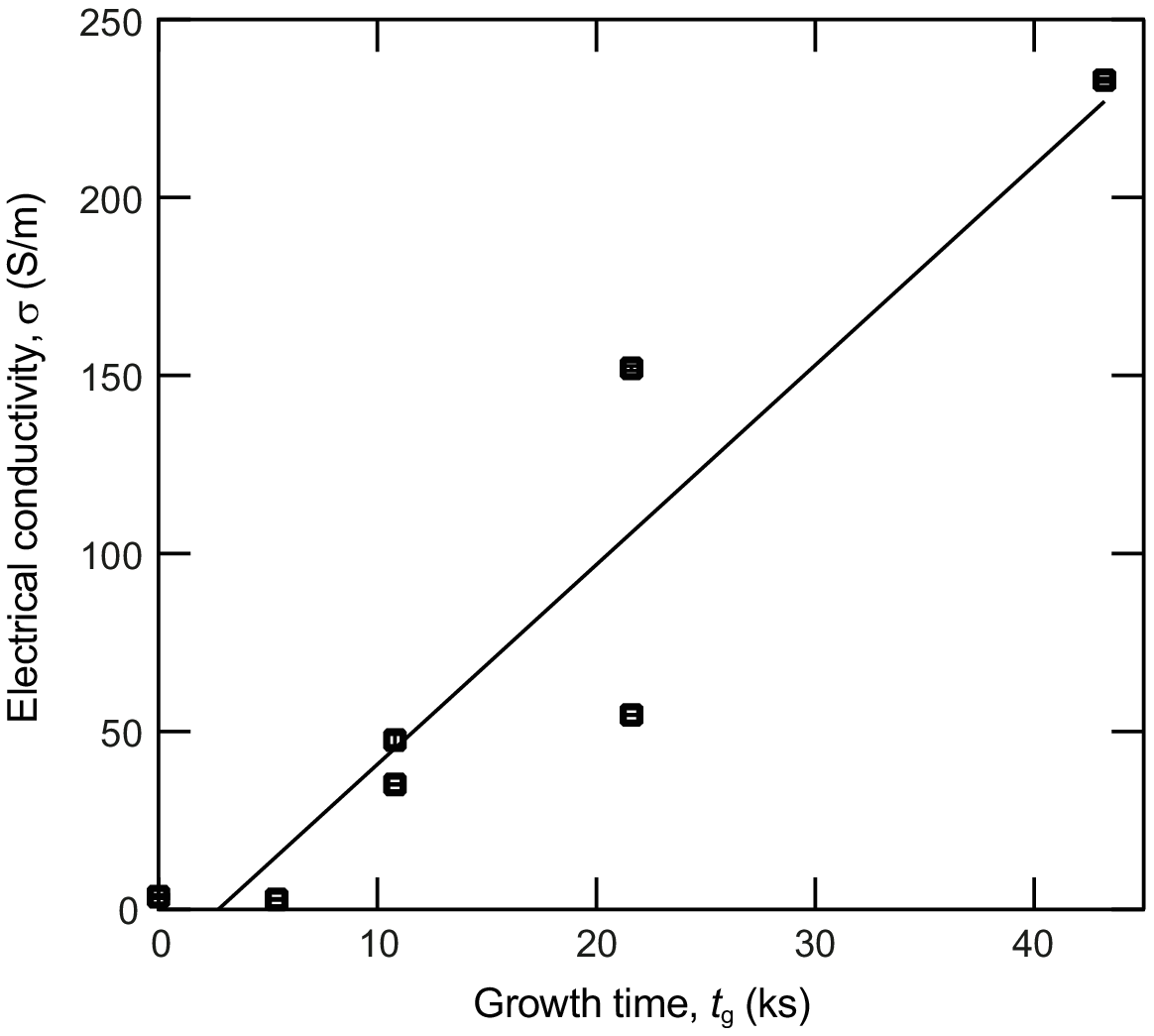}
\caption{Examples of nanopillared and bare membrane temperature in the center of the heated spot as a function of absorbed UV laser power.~(a) Set A and (b) Set B. }
\label{fig:FigS7}
\end{center}
\end{figure}

Examples of the Raman thermometry data are shown in Fig.~\ref{fig:FigS7}, where we plot Raman-derived temperature at the center of the heated spot as a function of absorbed laser power for representative specimens with nanopillars and their corresponding control membranes without nanopillars.~The Raman thermometry data show predominantly linear temperature increases for absorbed powers up to $\sim$2.0 mW. In order to determine the ratio of thermal conductivities described in Eq.~(1), we have carried out linear fits to the Raman thermometry data. For highest accuracy, we fit data points in the range of 40 $^{\circ}$C to 120 $^{\circ}$C.~We exclude lower temperatures because the Raman peak shifts are small and the corresponding uncertainty is high for these points.~For higher temperatures, the response starts to become nonlinear due to the decrease in the thermal conductivity of the Si membrane with increasing temperature.~Data for multiple membranes from the same run were fit individually and then their slopes $b$ were averaged using a weighting factor of $\sigma_{b}/b$, where $\sigma_{b}$ is the standard deviation of the set of measured slopes.~The resulting ratios are shown in Table~\ref{tab:table1} with corresponding uncertainties.~As portrayed in the table, $k_{\textup{NPM}}/k_{\textup{Mem}}$ values are predominantly less than unity with a minimum value of 0.79, i.e., a 21~$\%$ reduction in the thermal conductivity.~We note here that run number D442 was a control sample with only an AlN buffer layer on the Si membrane. It exhibits a thermal conductivity that is nearly identical to the pristine membrane.~As discussed in the main article, partial coalescence of the nanopillars reduces the NPM effect and leads to higher thermal conductivities. 

\section{Possible additional heat transport mechanisms}

The presence of the nanopillar roots on the membranes may be viewed as a source of surface roughness, and thus might be thought to increase surface/boundary scattering of phonons and thereby reduce the thermal conductivity, similar to investigations of rough Si nanowires~\cite{Hochbaum2008,Donadio_2009,He_2012} and rough Si membranes~\cite{neogi2015tuning}.~However, the average width of the nanopillar roots among the various fabrication runs in our investigation is 65 nm (see Table~\ref{tab:table2}).~This is significantly larger than the range of phonon wavelengths ($\sim$5 to $\sim$60 angstroms for Si at room temperature~\cite{Esfarjani2011}), which is the size scale relevant for boundary scattering due to roughness~\cite{Hochbaum2008,neogi2015tuning}.~Furthermore, if boundary scattering of any form is dominant\footnote{For example, in Ref.~\cite{Maire_2018} the presence of nanopillars was interpreted as a source of increased boundary scattering.}, the thermal conductivity would not increase with an increase in nanopillar coalescence occurring predominantly at the tips; yet as noted in the main article we observe a significant rise in $k$ in the severely coalesced case of 5.7~$\mu$m tall nanopillars (see Fig.~3a).

Secondly, it is conceivable that a portion of the heat might take a path through the coalesced nanopillars which would effectively serve as a parallel ``branch" for the thermal transport for those cases, resulting in a larger overall area through which the heat-carrying phonons travel.~While this is unlikely because the phonons will scatter at the coalesced nanopillar tips, the increased overall cross-sectional area implies lower $k$ for the same power transmitted$-$thus improving the reduction rather than deteriorating it as we observe for the 5.7~$\mu$m tall nanopillars in Fig.~3.

Furthermore, we note that any transport of heat to the environment rather than being delivered to the substrate at the edges of either type of membrane (i.e., base membrane in an NPM or bare membrane) would result in an overestimation of the integrated thermal power $P$ flowing through the membrane, which we assumed to be equal to $P_{\rm{abs}}$.~One example of such transport would be heat loss through the small but finite thermal conductivity of the surrounding air.~We here show that any such mechanism results in an~\textit{overestimation} of the true thermal conductivity $k^{\textup{true}}$ of the membrane.~Postulating that this mechanism results in $P = \eta P_{\textup{abs}}$, where $\eta$ is a number less than 1, and defining $ D = 2 \pi d /(\gamma/2 +\textup{ln}(R/r_0))$, we can combine Eqs.~\ref{equation:eqnS1} and \ref{equation:eqnS4} and solve for $k^{\textup{true}}$ as
\begin{equation}
\begin{split}
    k^{\textup{true}} &= \frac{1}{D} \left[ \frac{\partial \Delta T(0)}{\partial (\eta P_{\textup{abs}})}\right]^{-1}= \frac{1}{D} \left[\frac{1}{\eta}\frac{\partial \Delta T(0)}{ \partial P_{\textup{abs}}}\right]^{-1}\\
    &= \eta \frac{1}{D} \left[\frac{\partial \Delta T(0)}{ \partial P_{\textup{abs}}}\right]^{-1} = \eta  k^{\textup{calc}} ,\\
\end{split}
\end{equation}
where $k^{\textup{calc}}$ is the value we would calculate for $k$ using our original data analysis method.~Because $\eta$ is less than 1, the true value $k^{\textup{true}}$ would be less than our calculated value $k^{\textup{calc}}$.\\
\indent In taking the ratio of $k_{\textup{NPM}}/k_{\textup{Mem}}$, any heat conduction through the surrounding air would affect both the NPM membrane and the bare membrane, and therefore the heat loss would cancel in taking this ratio, provided the nanopillars do not affect local interactions with the air.~Although the nanopillars bear a superficial resemblance to heat sink structures in which fins are used to conduct heat away from an object and increase the surface area subject to air cooling, the spacing between the nanopillars on our membranes is smaller than or comparable to the mean free path of air molecules at atmospheric pressure, around 65 nm~\cite{Jennings1988}.~This tight spacing precludes the development of natural convective or conductive regions in the gaps between nanopillars or any effective increase in membrane surface area by the nanopillars. Instead, the high thermal conductivity along the axis of the nanopillars (compared with air) effectively moves the membrane/air interface to the plane defined by the tips of the nanopillars.~However, even if the nanopillars were to increase air cooling to some small degree, the $\eta$ value for NPM membranes would then be smaller than that for the bare membranes, and thus the true ratio of thermal conductivities would be smaller than our measured ratio, i.e., we would be \textit{underestimating} the NPM effect. This calculation can also be understood qualitatively in that any loss of heat to the environment reduces the temperature difference generated across the path of electrical current for a given thermal power input, and therefore lowers the thermoelectric performance of a structure.  

\section{Dopant diffusion into the Si membrane}
\begin{figure} [!t]
\begin{center}\includegraphics[scale=0.75]{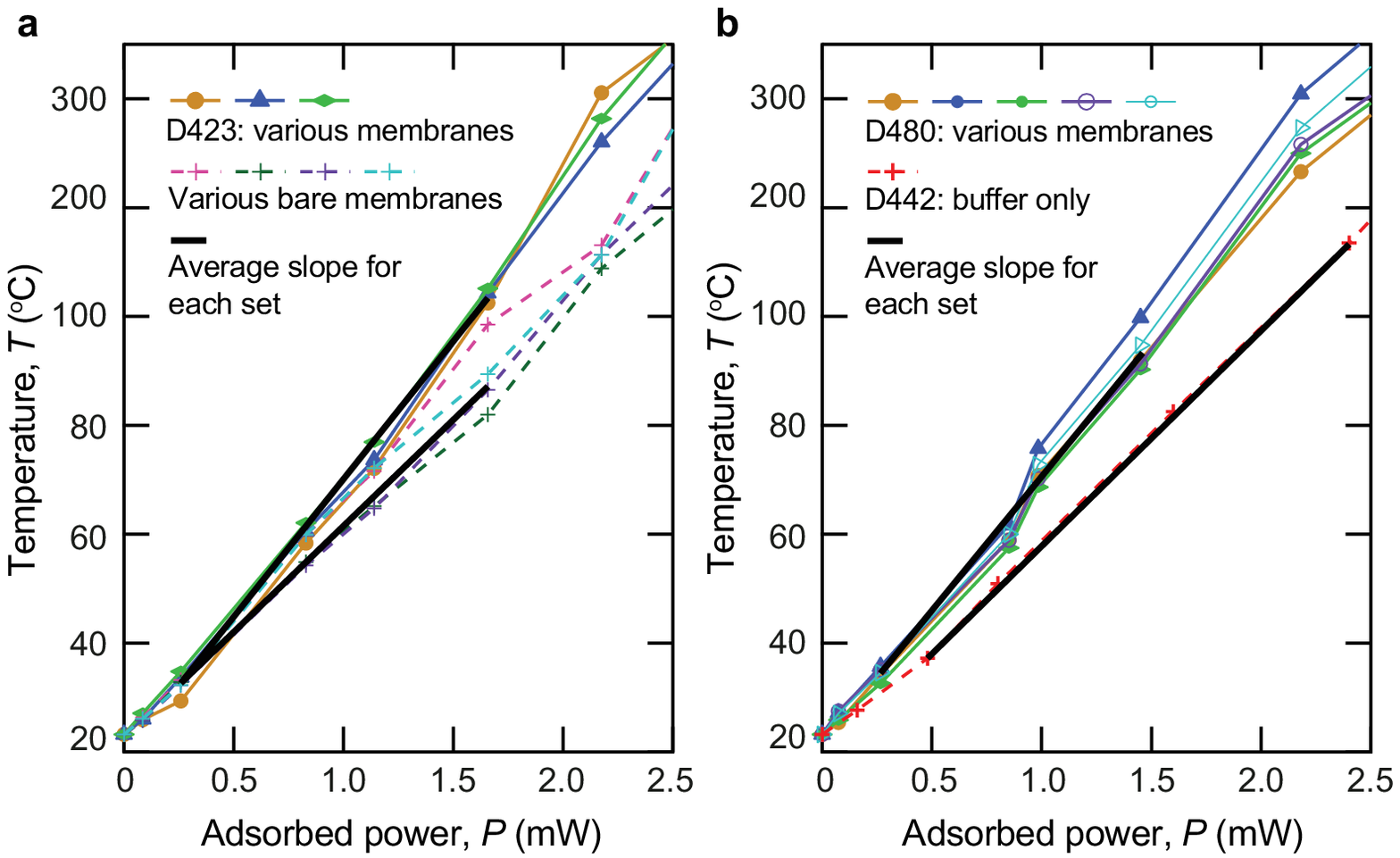}
\caption{Electrical conductivity of specimen Set A showing a linear increase as a function of nanopillar growth time.~The change is due to diffusion of Ga into the Si membrane.}
\label{fig:FigS8}
\end{center}
\end{figure}

Al and Ga are both p-type dopants in silicon, and will diffuse into to the thin Si membrane at high temperature, producing changes in membrane carrier concentration and electrical conductivity that depend on time at high temperature and available dopant concentrations.~For the specimen set without buffer layers, Set A, Ga flux is continually arriving at the Si membrane surface between the roots of the nanopillars.~From the measured electrical conductivity values, the Ga concentration in the membrane reaches a maximum value of 4$\times$10$^{16}$ cm$^{-3}$ for the specimen with the longest growth time, D423.~We infer the carrier concentration from typical relationships between electrical conductivity and carrier concentration for Si~\cite{Sze1981}.~As shown in Fig.~\ref{fig:FigS8}, the increase in $\sigma$ is linearly dependent on growth time.~This dependence implies that the Ga concentration and hence the hole concentration is a function of the total flux impinging on the top surface during NPM growth.~The maximum carrier concentration corresponds to a Ga concentration in the membrane of less than 1 ppm.~Figure~\ref{fig:FigS8} also illustrates that there was significant spatial variation in the measured electrical conductivity. For most of the runs, two areas on the die were sampled for the electrical conductivity, and although each measurement had low uncertainty, the spatial variations are on the order of 50 $\%$ to 150 $\%$ for the specimens with the highest $\sigma$.~The actual Ga flux to the wafer is likely much more uniform that this, but the local environment of randomly growing nanopillars would lead to large variations in shadowing of the surface by the nanopillars.\\
\indent In order to avoid the complication of diffusion-induced electrical conductivity changes within a specimen set, Set B employed an 8 nm thick AlN buffer layer as a diffusion blocking layer.~AlN has been used to block diffusion of Mg in GaN/AlGaN structures~\cite{Chowdhury2011} and of Al in TiN contact layers~\cite{Chun2001}.~As shown in the main article, this very thin buffer layer achieved our goal of removing the dependence of electrical conductivity on growth time.~The electrical conductivity for the specimen with the AlN buffer layer only (and no nanopillars) is still significantly higher than that measured for a pristine Si membrane, around 170 S/m vs. 4 S/m, and we attribute this increase to Al diffusion during the buffer layer growth.~Al diffusion out of AlN buffer layers is a contributing factor to conductive loss in RF devices grown on GaN-on-Si substrates~\cite{Bah2020}.~Al diffuses more readily in Si than Ga~\cite{Krause2002, Makris1971, okamura1971}, but Al flux was only present during the AlN buffer layer growth, lasting 180 s.~The AlN buffer did appear to be effective as a barrier to subsequent Ga diffusion based on the conductivity of the specimen with the longest growth time, D481.~For this run, the electrical conductivity would have more than doubled if Ga diffusion occurred at the same rate as in Set A, but in fact it remained unchanged.  

This diffusion is not a fundamental limitation for practical applications because it is only significant when the starting materials have very low background carrier concentrations.~SOI with 200 nm device layers are only available with low B doping concentrations ($<$1$\times$10$^{15}$ cm$^{-3}$) due to their method of fabrication. For practical thermoelectric applications, much higher electrical conductivity would be needed, and doping methods would be developed for the membranes to increase the electrical conductivity and therefore increase $ZT$. 

\bibliographystyle{ieeetr}
\bibliography{reflist}

\begin{thebibliography}{10}

\bibitem{Rowe1995}
D.~M. Rowe, {\em CRC Handbook of Thermoelectrics}.
\newblock CRC/Taylor and Francis, Boca Raton, 1995.

\bibitem{Yan2021}
Q.~Yan and M.~G. Kanatzidis, ``High-performance thermoelectrics and challenges
  for practical devices,'' {\em Nat. Mater.}, vol.~21, pp.~503--513, 2021.

\bibitem{Seebeck1826}
T.~J. Seebeck, ``Ueber die magnetische polarisation der metalle und erze durch
  temperatur-differenz,'' {\em Ann. Phys.}, vol.~82, pp.~253--286, 1826.

\bibitem{Peltier1834}
J.~C.~A. Peltier, ``Nouvelles expériences sur la caloricité des courants
  électrique,'' {\em Ann. Chim. Phys.}, vol.~56, pp.~371--386, 1834.

\bibitem{Beretta2019}
D.~Beretta, N.~Neophytou, J.~M. Hodges, M.~G. Kanatzidis, D.~Narducci,
  M.~Martin-Gonzalez, M.~Beekman, B.~Balke, G.~Cerretti, W.~Tremel,
  A.~Zevalkink, A.~I. Hofmann, C.~Muller, B.~Dorling, M.~Campoy-Quiles, and
  M.~Caironi, ``Thermoelectrics: From history, a window to the future,'' {\em
  Mat. Sci. Eng. R}, vol.~138, pp.~210--255, 2019.

\bibitem{Balandin_1998}
A.~Balandin and K.~L. Wang, ``Significant decrease of the lattice thermal
  conductivity due to phonon confinement in a free-standing semiconductor
  quantum well,'' {\em Phys. Rev. B}, vol.~58, pp.~1544--1549, 1998.

\bibitem{Berman1978}
R.~Berman, {\em Thermal Conduction in Solids}.
\newblock Clarendon Press, Oxford, UK, 1978.

\bibitem{Donadio_2009}
D.~Donadio and G.~Galli, ``Atomistic simulations of heat transport in silicon
  nanowires,'' {\em Phys. Rev. Lett.}, vol.~102, p.~195901, 2009.

\bibitem{He_2012}
Y.~He and G.~Galli, ``Microscopic origin of the reduced thermal conductivity of
  silicon nanowires,'' {\em Phys. Rev. Lett.}, vol.~108, p.~215901, 2012.

\bibitem{neogi2015tuning}
S.~Neogi, J.~S. Reparaz, L.~F.~C. Pereira, B.~Graczykowski, M.~R. Wagner,
  M.~Sledzinska, A.~Shchepetov, M.~Prunnila, J.~Ahopelto, C.~M.
  Sotomayor-Torres, {\em et~al.}, ``Tuning thermal transport in ultrathin
  silicon membranes by surface nanoscale engineering,'' {\em ACS Nano}, vol.~9,
  pp.~3820--3828, 2015.

\bibitem{Biswas_2012}
K.~Biswas, J.~He, I.~D. Blum, T.~P. Wu, C.-I.~Hogan, D.~N. Seidman, V.~P.
  Dravid, and M.~G. Kanatzidis, ``High-performance bulk thermoelectrics with
  all-scale hierarchical architectures,'' {\em Nature}, vol.~489, pp.~414--418,
  2012.

\bibitem{Snyder2008}
G.~J. Snyder and E.~S. Toberer, ``Complex thermoelectric materials,'' {\em Nat.
  Mater.}, vol.~7, no.~2, pp.~105--114, 2008.

\bibitem{Vineis2010}
C.~J. Vineis, A.~Shakouri, A.~Majumdar, and M.~G. Kanatzidis, ``Nanostructured
  thermoelectrics: Big efficiency gains from small features,'' {\em Adv.
  Mater.}, vol.~22, pp.~3970--3980, 2010.

\bibitem{Hochbaum2008}
A.~I. Hochbaum, R.~K. Chen, R.~D. Delgado, W.~J. Liang, E.~C. Garnett,
  M.~Najarian, A.~Majumdar, and P.~D. Yang, ``Enhanced thermoelectric
  performance of rough silicon nanowires,'' {\em Nature}, vol.~451,
  pp.~163--167, 2008.

\bibitem{Boukai2008}
A.~I. Boukai, Y.~Bunimovich, J.~Tahir-Kheli, J.~K. Yu, W.~A. Goddard, and J.~R.
  Heath, ``Silicon nanowires as efficient thermoelectric materials,'' {\em
  Nature}, vol.~451, pp.~168--171, 2008.

\bibitem{Esfarjani2011}
K.~Esfarjani, G.~Chen, and H.~T. Stokes, ``Heat transport in silicon from
  first-principles calculations,'' {\em Phys. Rev. B}, vol.~84, p.~085204,
  2011.

\bibitem{JuGoodson1999}
Y.~S. Ju and K.~E. Goodson, ``Phonon scattering in silicon films with thickness
  of order 100 nm,'' {\em Appl. Phys. Lett.}, vol.~74, pp.~3005--3007, 1999.

\bibitem{Garg_2018}
J.~Garg, T.~Luo, and G.~Chen, ``Spectral concentration of thermal conductivity
  in gan—a first-principles study,'' {\em Appl. Phys. Lett.}, vol.~112,
  p.~252101, 2018.

\bibitem{Ohishi2015}
Y.~Ohishi, J.~Xie, Y.~Miyazaki, Y.~Aikebaier, H.~Muta, K.~Kurosaki,
  S.~Yamanaka, N.~Uchida, and T.~Tada, ``Thermoelectric properties of heavily
  boron- and phosphorus-doped silicon,'' {\em Jpn. J. Appl. Phys.}, vol.~54,
  p.~071301, 2015.

\bibitem{Kim2015}
H.-S. Kim, Z.~M. Gibbs, Y.~Tang, H.~Wang, and G.~J. Snyder, ``Characterization
  of lorenz number with seebeck coefficient measurement,'' {\em APL Materials},
  vol.~3, p.~041506, 2015.

\bibitem{DavisHussein2014}
B.~L. Davis and M.~I. Hussein, ``Nanophononic metamaterial: Thermal
  conductivity reduction by local resonance,'' {\em Phys. Rev. Lett.},
  vol.~112, p.~055505, 2014.

\bibitem{Wei2015}
Z.~Y. Wei, J.~K. Yang, K.~D. Bi, and Y.~F. Chen, ``Phonon transport properties
  in pillared silicon film,'' {\em J. Appl. Phys.}, vol.~118, p.~155103, 2015.

\bibitem{Honarvar2016a}
H.~Honarvar and M.~I. Hussein, ``Spectral energy analysis of locally resonant
  nanophononic metamaterials by molecular simulations,'' {\em Phys. Rev. B},
  vol.~93, p.~081412(R), 2016.

\bibitem{Xiong2016}
S.~Y. Xiong, K.~Saaskilahti, Y.~A. Kosevich, H.~X. Han, D.~Donadio, and
  S.~Volz, ``Blocking phonon transport by structural resonances in alloy-based
  nanophononic metamaterials leads to ultralow thermal conductivity,'' {\em
  Phys. Rev. Lett.}, vol.~117, p.~025503, 2016.

\bibitem{Honarvar2018}
H.~Honarvar and M.~I. Hussein, ``Two orders of magnitude reduction in silicon
  membrane thermal conductivity by resonance hybridizations,'' {\em Phys. Rev.
  B}, vol.~97, p.~195413, 2018.

\bibitem{Hussein_2018}
M.~I. Hussein and H.~Honarvar, {\em Handbook of Materials Modeling:
  Applications: Current and Emerging Materials (Eds: W. Andreoni, S. Yip), Vol.
  2, pp. 1–21}, ch.~17-1: Resonant thermal transport in nanophononic
  metamaterials.
\newblock Springer, New York, 2018.

\bibitem{Hussein2020}
M.~I. Hussein, C.~N. Tsai, and H.~Honarvar, ``Thermal conductivity reduction in
  a nanophononic metamaterial versus a nanophononic crystal: A review and
  comparative analysis,'' {\em Adv. Funct. Mater.}, vol.~30, p.~1906718, 2020.

\bibitem{Tang2010}
J.~Tang, H.-T. Wang, D.~H. Lee, M.~Fardy, Z.~Huo, T.~P. Russell, and P.~Yang,
  ``Holey silicon as an efficient thermoelectric material,'' {\em Nano Lett.},
  vol.~10, pp.~4279--4283, 2010.

\bibitem{Yu2010}
J.~K. Yu, S.~Mitrovic, D.~Tham, J.~Varghese, and J.~R. Heath, ``Reduction of
  thermal conductivity in phononic nanomesh structures,'' {\em Nat.
  Nanotechnol.}, vol.~5, pp.~718--721, 2010.

\bibitem{bertness2006}
K.~A. Bertness, A.~Roshko, N.~A. Sanford, J.~A. Barker, and A.~V. Davydov,
  ``Spontaneously grown {GaN} and {AlGaN} nanowires,'' {\em J. Cryst. Growth},
  vol.~287, pp.~522--527, 2006.

\bibitem{Cai2010}
W.~Cai, A.~L. Moore, Y.~Zhu, X.~Li, S.~Chen, L.~Shi, and R.~S. Ruoff, ``Thermal
  transport in suspended and supported monolayer graphene grown by chemical
  vapor deposition,'' {\em Nano Lett.}, vol.~10, p.~1645, 2010.

\bibitem{Luo2014}
Z.~Luo, H.~Liu, B.~T. Spann, Y.~Feng, P.~Ye, Y.~P. Chen, and X.~Xu,
  ``Measurement of in-plane thermal conductivity of ultrathin films using
  micro-{R}aman spectroscopy,'' {\em Nanosc. Microsc.Therm.}, vol.~18,
  pp.~183--193, 2014.

\bibitem{luo2015}
Z.~Luo, J.~Maassen, Y.~Deng, Y.~Du, R.~P. Garrelts, M.~S. Lundstrom, D.~Y.
  Peide, and X.~Xu, ``Anisotropic in-plane thermal conductivity observed in
  few-layer black phosphorus,'' {\em Nat. Commun.}, vol.~6, p.~8572, 2015.

\bibitem{chavez2014}
E.~Ch{\'a}vez-{\'A}ngel, J.~Reparaz, J.~Gomis-Bresco, M.~Wagner, J.~Cuffe,
  B.~Graczykowski, A.~Shchepetov, H.~Jiang, M.~Prunnila, J.~Ahopelto, {\em
  et~al.}, ``Reduction of the thermal conductivity in free-standing silicon
  nano-membranes investigated by non-invasive {R}aman thermometry,'' {\em APL
  Mater.}, vol.~2, p.~012113, 2014.

\bibitem{cuffe2012}
J.~Cuffe, E.~Chavez, A.~Shchepetov, P.~O. Chapuis, E.~H. El~Boudouti,
  F.~Alzina, T.~Kehoe, J.~Gomis-Bresco, D.~Dudek, Y.~Pennec,
  B.~Djafari-Rouhani, M.~Prunnila, J.~Ahopelto, and C.~M.~S. Torres, ``Phonons
  in slow motion: Dispersion relations in ultrathin {Si} membranes,'' {\em Nano
  Lett.}, vol.~12, pp.~3569--3573, 2012.

\bibitem{wagner2007}
M.~Wagner, {\em Simulation of thermoelectric devices}.
\newblock Doctoral Dissertation, Technical University of Vienna, Austria, 2007.

\bibitem{Geballe1955}
T.~H. Geballe and G.~W. Hull, ``Seebeck effect in silicon,'' {\em Phys. Rev.},
  vol.~98, pp.~940--947, 1955.

\bibitem{Liu2000}
Z.~Liu, X.~Zhang, Y.~Mao, Y.~Y. Zhu, Z.~Yang, C.~T. Chan, and P.~Sheng,
  ``Locally resonant sonic materials,'' {\em Science}, vol.~289,
  pp.~1734--1736, 2000.

\bibitem{Yang2020}
L.~Yang, Z.-G. Chen, M.~S. Dargusch, and J.~Zou, ``High performance
  thermoelectric materials: Progress and their applications,'' {\em Adv. Energy
  Mater.}, vol.~8, p.~1701797, 2018.

\bibitem{weber2017}
J.~C. Weber, M.~D. Brubaker, T.~M. Wallis, and K.~A. Bertness, ``Lithographic
  sonication patterning of large area {G}a{N} nanopillar forests grown on a
  {S}i substrate,'' {\em Microelectron. Eng.}, vol.~181, pp.~43--46, 2017.

\bibitem{reparaz2014}
J.~Reparaz, E.~Chavez-Angel, M.~Wagner, B.~Graczykowski, J.~Gomis-Bresco,
  F.~Alzina, and C.~Sotomayor~Torres, ``A novel contactless technique for
  thermal field mapping and thermal conductivity determination: Two-laser
  {R}aman thermometry,'' {\em Rev. Sci. Instrum.}, vol.~85, p.~034901, 2014.

\bibitem{cardona1984}
J.~Men\'endez and M.~Cardona, ``Temperature dependence of the first-order raman
  scattering by phonons in {Si}, {Ge}, and
  $\ensuremath{\alpha}\ensuremath{-}\mathrm{S}\mathrm{n}$: Anharmonic
  effects,'' {\em Phys. Rev. B}, vol.~29, pp.~2051--2059, 1984.

\bibitem{Tersoff1988}
J.~Tersoff, ``Empirical interatomic potential for silicon with improved elastic
  properties,'' {\em Phys. Rev. B}, vol.~38, pp.~9902--9905, 1988.

\bibitem{Nord2003}
J.~Nord, K.~Albe, P.~Erhart, and K.~Nordlund, ``Modelling of compound
  semiconductors: analytical bond-order potential for gallium, nitrogen and
  gallium nitride,'' {\em J. Phys.-Condens Mat.}, vol.~15, pp.~5649--5662,
  2003.

\bibitem{Tersoff1989}
J.~Tersoff, ``Modeling solid-state chemistry: Interatomic potentials for
  multicomponent systems,'' {\em Phys. Rev. B}, vol.~39, pp.~5566--5568, 1989.

\bibitem{Gale2003}
J.~Gale and A.~Rohl, ``The general utility lattice program (gulp),'' {\em Mol.
  Simul.}, vol.~29, pp.~291--341, 2003.

\bibitem{schelling2002comparison}
P.~K. Schelling, S.~R. Phillpot, and P.~Keblinski, ``Comparison of atomic-level
  simulation methods for computing thermal conductivity,'' {\em Phys. Rev. B},
  vol.~65, p.~144306, 2002.

\bibitem{plimpton1995fast}
S.~Plimpton, ``Fast parallel algorithms for short-range molecular dynamics,''
  {\em J. Comput. Phys.}, vol.~117, pp.~1--19, 1995.

\bibitem{bertness2014}
K.~A. Bertness, M.~D. Brubaker, T.~E. Harvey, S.~M. Duff, A.~W. Sanders, and
  N.~A. Sanford, ``In situ temperature measurements for selective epitaxy of
  gan nanowires,'' {\em Phys. Status Solidi C}, vol.~11, no.~3-4, pp.~590--593,
  2014.

\bibitem{brubaker2016}
M.~D. Brubaker, S.~M. Duff, T.~E. Harvey, P.~T. Blanchard, A.~Roshko, A.~W.
  Sanders, N.~A. Sanford, and K.~A. Bertness, ``Polarity-controlled {GaN}/{AlN}
  nucleation layers for selective-area growth of gan nanowire arrays on si
  (111) substrates by molecular beam epitaxy,'' {\em Cryst. Growth Des.},
  vol.~16, no.~2, pp.~596--604, 2016.

\bibitem{ExpInt}
N.~I. of~Standards and Technology, ``Digital library of mathematical
  functions.''

\bibitem{Harris1957}
F.~Harris, ``Tables of the exponential integral ei(x),'' {\em Math. Comput.},
  vol.~11, p.~9, 1957.

\bibitem{Maire_2018}
J.~Maire, R.~Anufriev, T.~Hori, J.~Shiomi, S.~Volz, and M.~Nomura, ``Thermal
  conductivity reduction in silicon fishbone nanowires,'' {\em Sci. Rep.},
  vol.~8, p.~4452, 2018.

\bibitem{Jennings1988}
S.~G. Jennings, ``The mean free path in air,'' {\em J. Aerosel Science},
  vol.~19, pp.~159--166, 1988.

\bibitem{Sze1981}
S.~M. Sze, {\em Physics of Semiconductor Devices, p. 32}.
\newblock John Wiley and Sons, Inc., New York, 1981.

\bibitem{Chowdhury2011}
S.~Chowdhury, B.~L. Swenson, J.~Lu, and U.~K. Mishra, ``Use of sub-nanometer
  thick aln to arrest diffusion of ion-implanted mg into regrown algan/gan
  layers,'' {\em Jpn. J. Appl. Phys.}, vol.~50, p.~101002, 2011.

\bibitem{Chun2001}
J.~S. Chun, C.~Desjardins, P.and~Lavoie, C.~S. Shin, C.~Cabral, I.~Petrov, and
  J.~E. Greene, ``Interfacial reactions in epitaxial al/tin(111) model
  diffusion barriers: Formation of an impervious self-limited
  wurtzite-structure aln(0001) blocking layer,'' {\em J. Appl. Phys.}, vol.~89,
  pp.~7841--7845, 2001.

\bibitem{Bah2020}
M.~Bah, D.~Valente, M.~Lesecq, N.~Defrance, M.~Garcia~Barros, J.~C. De~Jaeger,
  E.~Frayssinet, R.~Comyn, T.~H. Ngo, D.~Alquier, and Y.~Cordier, ``Electrical
  activity at the aln/si interface: identifying the main origin of propagation
  losses in gan-on-si devices at microwave frequencies,'' {\em Sci. Rep.},
  vol.~10, p.~14166, 2020.

\bibitem{Krause2002}
O.~Krause, H.~Ryssel, and P.~Pichler, ``Determination of aluminum diffusion
  parameters in silicon,'' {\em J. Appl. Phys.}, vol.~91, pp.~5645--5649, 2002.

\bibitem{Makris1971}
J.~S. Makris and B.~J. Masters, ``Gallium diffusions into silicon and
  boron‐doped silicon,'' {\em J. Appl. Phys.}, vol.~42, pp.~3750--3754, 1971.

\bibitem{okamura1971}
M.~Okamura, ``Gallium diffusion in silicon,'' {\em Jpn. J. Appl. Phys.},
  vol.~10, no.~4, p.~434, 1971.

\end{thebibliography}

\end{document}